\documentclass[11pt]{article}        

\usepackage{etoolbox}
\usepackage[dvipsnames]{xcolor}

\date{}

\usepackage{amsthm,amssymb}

\usepackage{graphicx}

\usepackage{amscd,amssymb}
\usepackage{amsmath}
\usepackage{tikz}
\usepackage{tikz-3dplot}
\usepackage{tikz-cd}
\usepackage{faktor}

\newtheorem  {theorem}                  {Theorem}
\newtheorem* {theorem*}                   {Theorem}

\newtheorem* {prop*}     {Proposition}

\theoremstyle{definition}
\newtheorem {defi}[theorem] {Definition}
\newtheorem {Remark} [theorem]         {Remark}

\newtheorem* {Example*}    {Example}

\newcommand{\pp}[2]{\frac{\partial#1}{\partial#2}}

\title{Local construction of knotted screw dislocations\\ in smectic liquid crystals}

\author{Robert Cardona and Andreu Vega}

\newcommand{\Addresses}{
{
  \bigskip

{\sc \noindent Robert Cardona}

\noindent Departament de Matem\`atiques i Inform\`atica, Universitat de Barcelona, Gran Via de Les Corts Catalanes 585, 08007 Barcelona, Spain; Centre de Recerca Matemàtica, Campus de Bellaterra, Edifici C, 08193, Barcelona, Spain.

{\noindent  \em robert.cardona@ub.edu\/}
}

  \medskip

{\sc \noindent Andreu Vega}

\noindent Departament de Matem\`atiques i Inform\`atica, Universitat de Barcelona, Gran Via de Les Corts Catalanes 585, 08007 Barcelona, Spain.

}

\begin{document}

\maketitle

\begin{abstract}
    The goal of this work is to show that, mathematically, screw dislocation loops with distinct topological properties can be introduced locally on a given smectic liquid crystal configuration. Concretely, given a configuration, we show that a new screw dislocation loop can be introduced along any knot or link transverse to the regular layers through a purely local modification. We define a topological invariant of screw dislocation loops, the multiplicity, and show that it can be explicitly prescribed in our construction. Finally, we apply this method to establish that any link type can be locally introduced within the set of screw dislocations of smectic configurations of a certain class. This shows that both the global topology (link type) and the local topology (multiplicity) of loop dislocations can be very rich, even when constrained to arise from local modifications.
\end{abstract}

\section{Introduction}
Liquid crystals are substances that have properties of both liquid and solid states of matter. One of their key crystalline properties is their orientational order: at each point, there is a preferred direction of the orientation of the molecules. At the same time, the molecules can flow like a liquid, with or without constraints. These constraints arise, for instance, in any phase of smectic liquid crystals admitting a layered structure reflecting positional order. Singularities, also known as defects, can occur in both the director and the layered structure of the liquid crystal, significantly influencing the material's physical properties \cite{FB}. In three-dimensional space, these defects manifest as points or lines. For smectic liquid crystals, line defects can be either disclinations or dislocations (of edge or screw type).\\

The first approach to study these defects is homotopy theory, which can be used to classify neighborhoods of defects up to homotopy \cite{Me}. One can also try to understand how line defects are globally positioned in the ambient space, as captured by knot theory. This ``global position" has been observed to be rather diverse in experiments \cite{KnotsExp}. Knot theory has thus been imported to the theoretical study of liquid crystals in various contexts \cite{MA, M1, NKTK}. In phases like smectics or cholesterics, the ordered medium has more structure than only a director line field, and other mathematical theories beyond homotopy theory and knot theory are involved. For instance, foliation theory is crucial in smectics \cite{Poe, CAK, MAHK, SeKa}, while contact topology found important applications in cholesterics \cite{M2, PA1, PoTh}.

In our discussion, we are interested in the knotted line dislocations of smectic liquid crystals, which will always be assumed to be in a phase with a layered structure. These involve singularities of the layered structure, which is modeled by a singular foliation, and their local classification involves not only homotopy theory but also the local behavior of the layers \cite{KM}. Two topological types of line dislocations are edge and screw defects, both of which can be observed in smectic crystals \cite{WK, RG}. The former entails a discontinuity in the layer spacing, while the latter consists of a helicoidal behavior of the layers around the singular line and occurs more frequently in nature. In the recent work of Severino, Kamien, and Bode \cite{SKB}, it is shown how to construct (theoretically) smectics with prescribed links, that is, collections of knotted closed curves with an arbitrary global position, in three dimensions. The construction is global: given a link, one produces (up to possibly adding a controlled number and type of additional singularities) a smectic that contains that given link as part of the line dislocations of the liquid crystal. The goal of this paper is to describe, mathematically, ways in which links and knots can be introduced into the set of dislocation singularities of a given smectic configuration through a \textit{local modification} (in a neighborhood of the knot). This is physically relevant, as it makes it more plausible for these loops to form naturally in smectics. The construction is general enough to exhibit a rich topological behavior both from a local and a global point of view, in a precise way that we will explain. \\

Before summarizing our results, let us point out that the core of our work is written in a mathematical language and consists of completely formal proofs that rely mostly on three-dimensional algebraic and geometric topology. However, in the expository Section \ref{s:informal}, we outline informally, mostly with figures, some of the ideas of the paper in a particular simple case. We refer to the classical textbooks \cite{Rol, Hat} for the reader interested in the background needed to understand the proofs in detail. \\

Our first results describe a way to introduce a link $L$, transverse to the layers of the smectic, to the set of screw loop defects of a smectic through a \textit{local procedure}. By this, we mean that the smectic is not modified away from a neighborhood of the link. In addition, we introduce an invariant of a smectic along a screw loop defect, the local multiplicity, and a global invariant of a given layer along a screw loop defect, the (global) multiplicity. Contrary to other invariants, like the twist of a screw defect \cite{SKB}, the local multiplicity is a topological invariant of the defect, while the global multiplicity is a topological invariant of a given layer. Our construction allows prescribing the value of these invariants along the transverse link $L$. We formalize the local modification of layers in Section \ref{s:local}, and introduce the invariants, along with an improved construction to prescribe them, in Section \ref{s:localmult}. Both constructions rely on the notion of ``helix box", introduced by Rechtman and the first author in \cite{CR}.

\noindent The local modification near a knot transverse to the layers that we describe produces two parallel copies of that link as a screw defect in our model. This aligns with the understanding that screw loop defects are unlikely to form in isolation. Instead, they form within a network of defects; for example, screw loop defects might pair with opposite helicoid orientations. This is sometimes referred to as ``nucleation of pairs of screw defects". We hope that our construction, in addition to formally proving that knotted and linked screw defects can form through a local procedure in smectics, provides some mathematical evidence for this phenomenon. With some exceptions like \cite{MKS}, the theoretical framework for this process was not thoroughly investigated in the literature, as pointed out in \cite{RG}.\\

The previous construction is then used to show that one can locally introduce any arbitrary link type as part of the set of screw dislocation of any starting smectic configuration within a certain class, namely, those that come from an open book layered structure in an arbitrary closed three-manifold (see Section \ref{s:back} for a precise definition). By our previous construction, this only requires giving an algorithm to deform an arbitrary link to one that is transverse to the regular layers of an open book smectic configuration, without changing the way in which it is knotted. We do it in Section \ref{s:link} following an idea of Becker \cite{Be}. Although we will work in a general setting, one can keep in mind the situation of most physical relevance: open book layered structures in the three-sphere. As described in \cite{SKB}, these layered structures admit projections to a ball and extensions to $\mathbb{R}^3$ to define smectic configurations that match the ground state away from the ball. In particular, our arguments show that for these configurations in $\mathbb{R}^3$, one can locally introduce any link, up to isotopy, by a local modification of the smectic.\\

Our construction opens up new directions for future research. For example, in our discussion, we have restricted attention to screw dislocations. It would be interesting to develop a local construction of knotted edge dislocations. This might require new ideas, such as finding knot representatives tangent to the regular layers or introducing point defects near the knot (see \cite[Section 4]{SKB}).

Another context where local (and even global) constructions of knotted defects are still lacking is that of cholesteric liquid crystals. Our construction can be adapted to cholesterics, where, recall, the pseudo-layered structure is not intrinsic but depends on the chosen direction along which the molecular orientation is measured as rotating. With our methods, one can build layered structures for a given cholesteric configuration that contain additional knotted screw dislocations. However, this seems less compelling, since such layered structures are more “twisted” and thus likely less energetically favorable. By contrast, constructing knotted disclination lines (rather than dislocations), either by global or local methods, appears much more promising and represents an interesting direction for future research.\\ 

\subsection*{Acknowledgments} 
We thank Benjamin Bode and Joseph Pollard for their helpful comments and suggestions. We are grateful to the anonymous referees, whose valuable suggestions improved the quality of this paper.

\noindent The first author acknowledges partial support from the AEI grant PID2023-147585NA-I00, the Departament de Recerca i Universitats de la Generalitat de Catalunya (2021 SGR 00697), and the Spanish State Research Agency, through the Severo Ochoa and María de Maeztu Program for Centers and Units of Excellence in R\&D (CEX2020-001084-M).

\section{An informal overview}\label{s:informal}

Before presenting formal definitions and proofs, we give an informal overview of the main constructions. We illustrate the ideas heuristically using a particularly simple example, referring mostly to figures to illustrate the ideas.\\

\begin{figure}[!ht]
    \centering
    \begin{tikzpicture}
    \node at (0,0) {\includegraphics[scale=0.3]{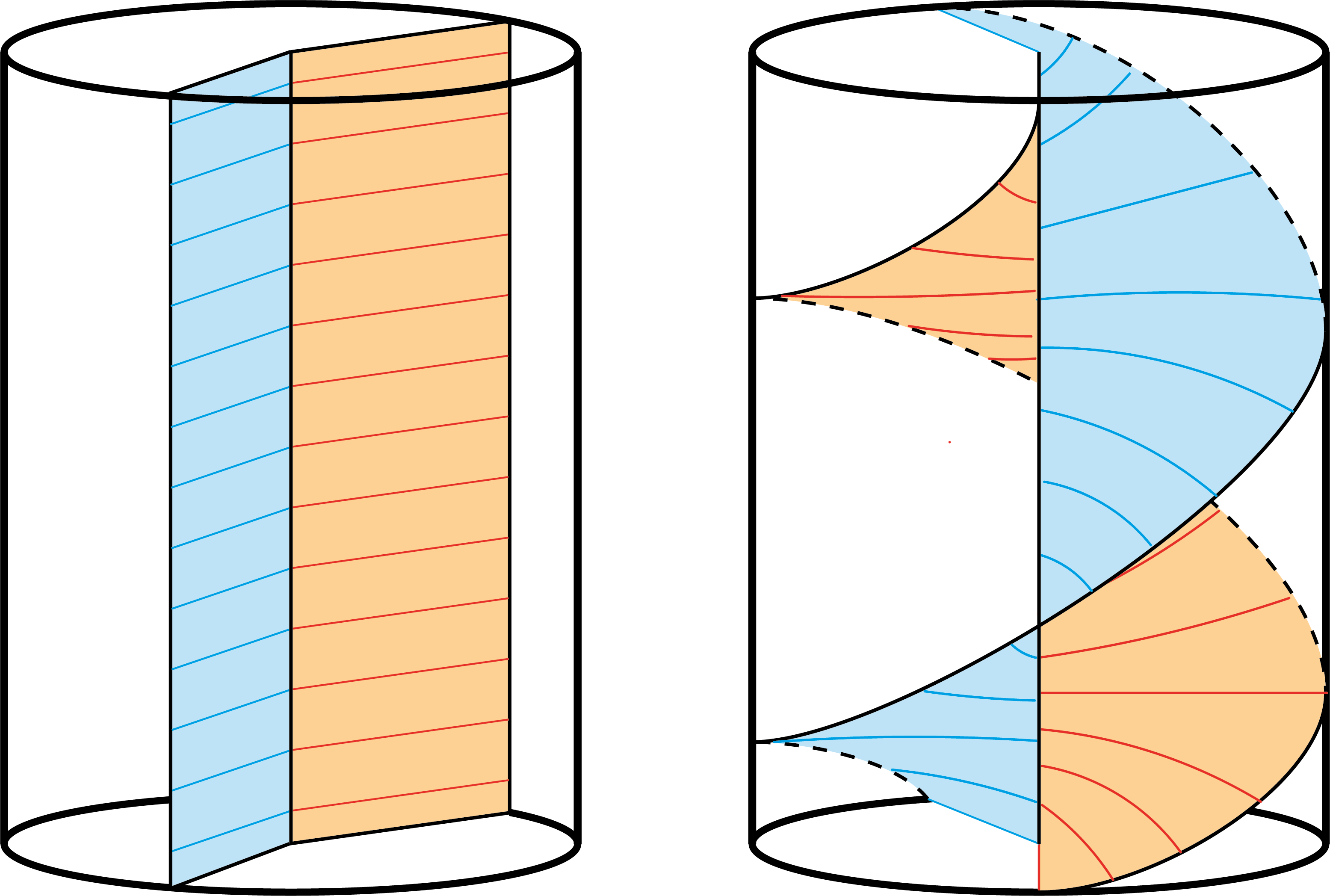}};
    \end{tikzpicture}
    \caption{Regular layers near a loop singularity}
    \label{fig:rational}
\end{figure}

Suppose that we are given a configuration of a smectic liquid crystal with a layered ordering. Mathematically, the layers are given by evenly spaced surfaces in a 3D space (a region of $\mathbb{R}^3$ or a compact manifold). These layers are well defined except perhaps along a set of points or line/loop singularities. These singularities, together with the ``regular" layers, partition the space. The left side of Figure \ref{fig:rational} illustrates two regular layers of an example of such a partition of a filled cylinder, given by ``radial surfaces". Every other regular layer is obtained by rotating one of these surfaces around the vertical axis. Hence, every point of the filled cylinder belongs to a regular layer except those points exactly on the $z$ axis, which is a line singularity. In this work, we will study a class of loop singularities called ``dislocations", which require the direction normal to the layers to behave nicely near the singularity. Near a loop singularity, if we identify a toroidal neighborhood of it with a cylinder whose top and bottom disks are identified by the identity map, the layers would have to come radially (as in the left side of Figure \ref{fig:rational}) or be helicoidal surfaces. The right side of Figure \ref{fig:rational} depicts an example of a helicoidal surface, but there are many other ways (that can be distinguished by a topological invariant, as we will see) in which these helicoidal surfaces can wrap around the loop singularity. 

The goal of this work is to show that new loop singularities can be introduced to the smectic configuration by modifying the layers only in a neighborhood of the new loop singularity. We think of the solid torus with radial surfaces (obtained from the cylinder above, identifying the top and bottom disks) as the layers of a starting smectic configuration.

\begin{figure}[!ht]
    \centering
    \begin{tikzpicture}
    \node at (0,0) {\includegraphics[scale=0.3]{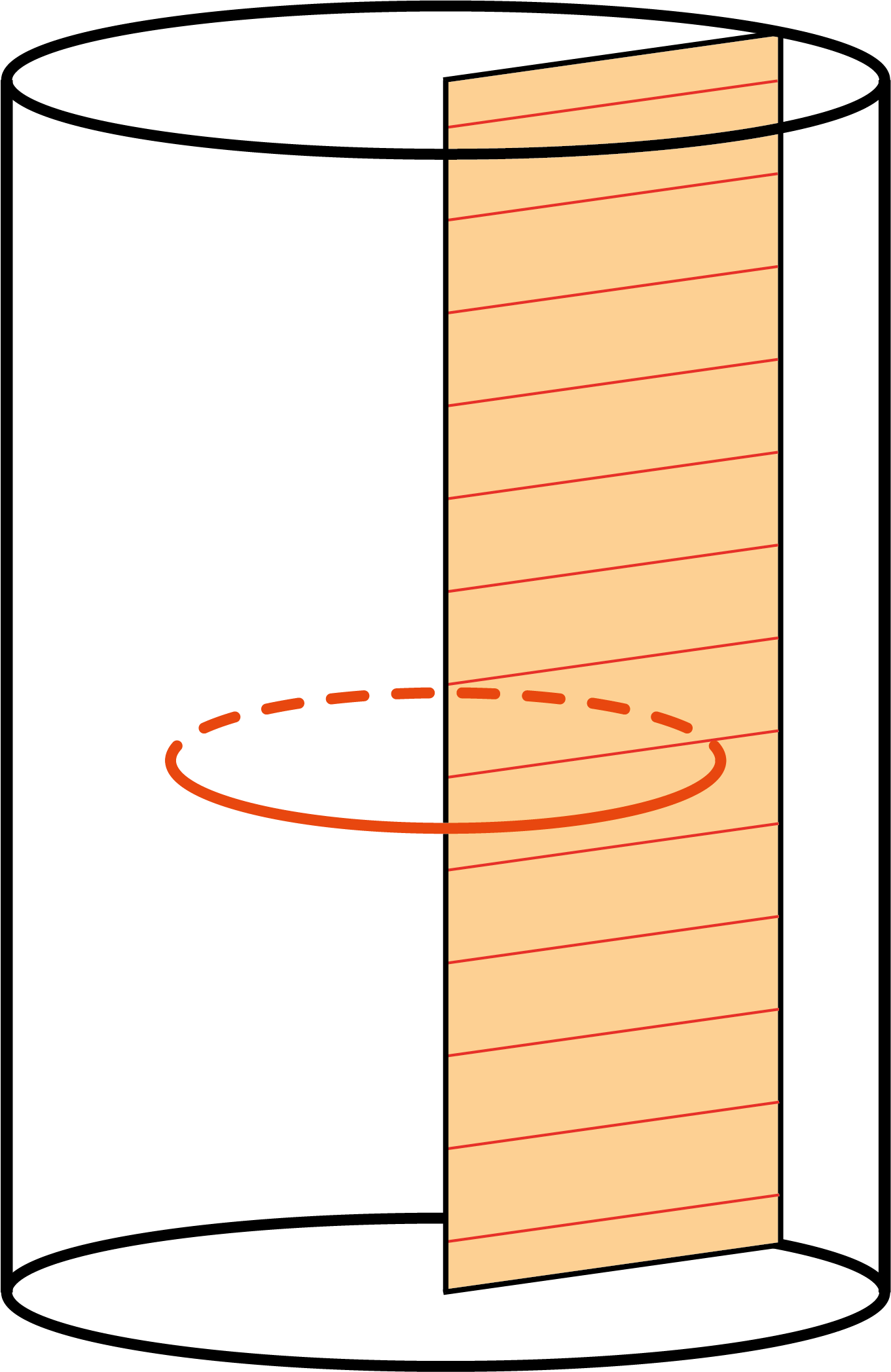}};
    \end{tikzpicture}
    \caption{A transverse unknot}
    \label{fig:unknot}
\end{figure}

For our first construction, we have to start with a knot that is initially transverse to the regular layers of the smectic configuration. We now consider a specific example to illustrate how a new dislocation loop can then be introduced. We will consider, for simplicity, an unknot transverse to the regular layers of the smectic configuration given by radial surfaces in the cylinder. This is depicted in Figure \ref{fig:unknot}. 

\begin{figure}[!ht]
    \centering
    \begin{tikzpicture}
    \node at (0,0) {\includegraphics[scale=0.3]{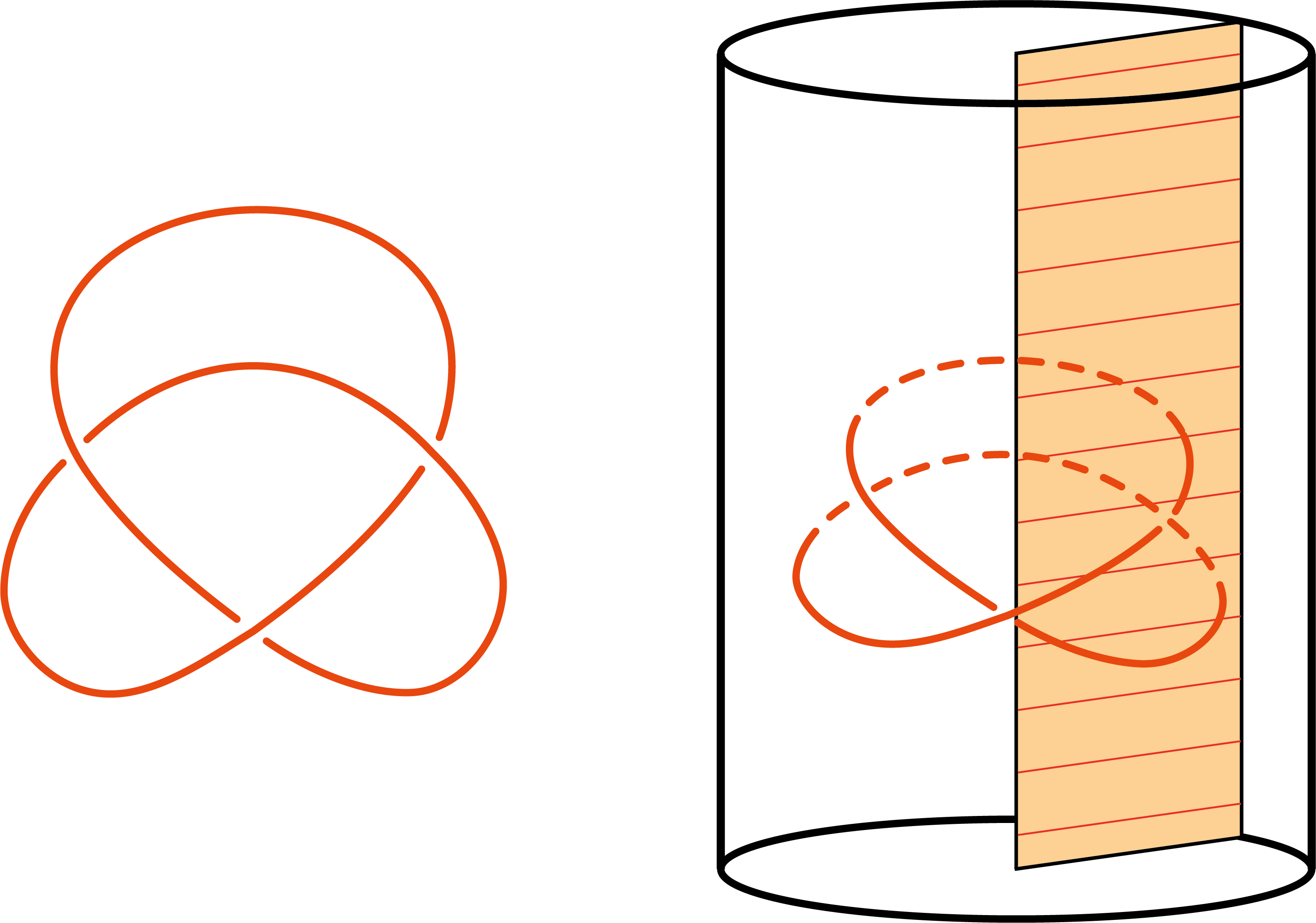}};
    \end{tikzpicture}
    \caption{A trefoil knot and a transverse representative}
    \label{fig:knot}
\end{figure}

Notice, however, that a priori any knot type could admit a representative that is transverse to the regular layers. For example Figure \ref{fig:knot} pictures the so-called ``trefoil knot", and another curve knotted in the same way but transverse to the layers of the smectic configuration given by radial surfaces in the cylinder.

\begin{figure}[!ht]
    \centering
    \begin{tikzpicture}
    \node at (0,0) {\includegraphics[scale=0.3]{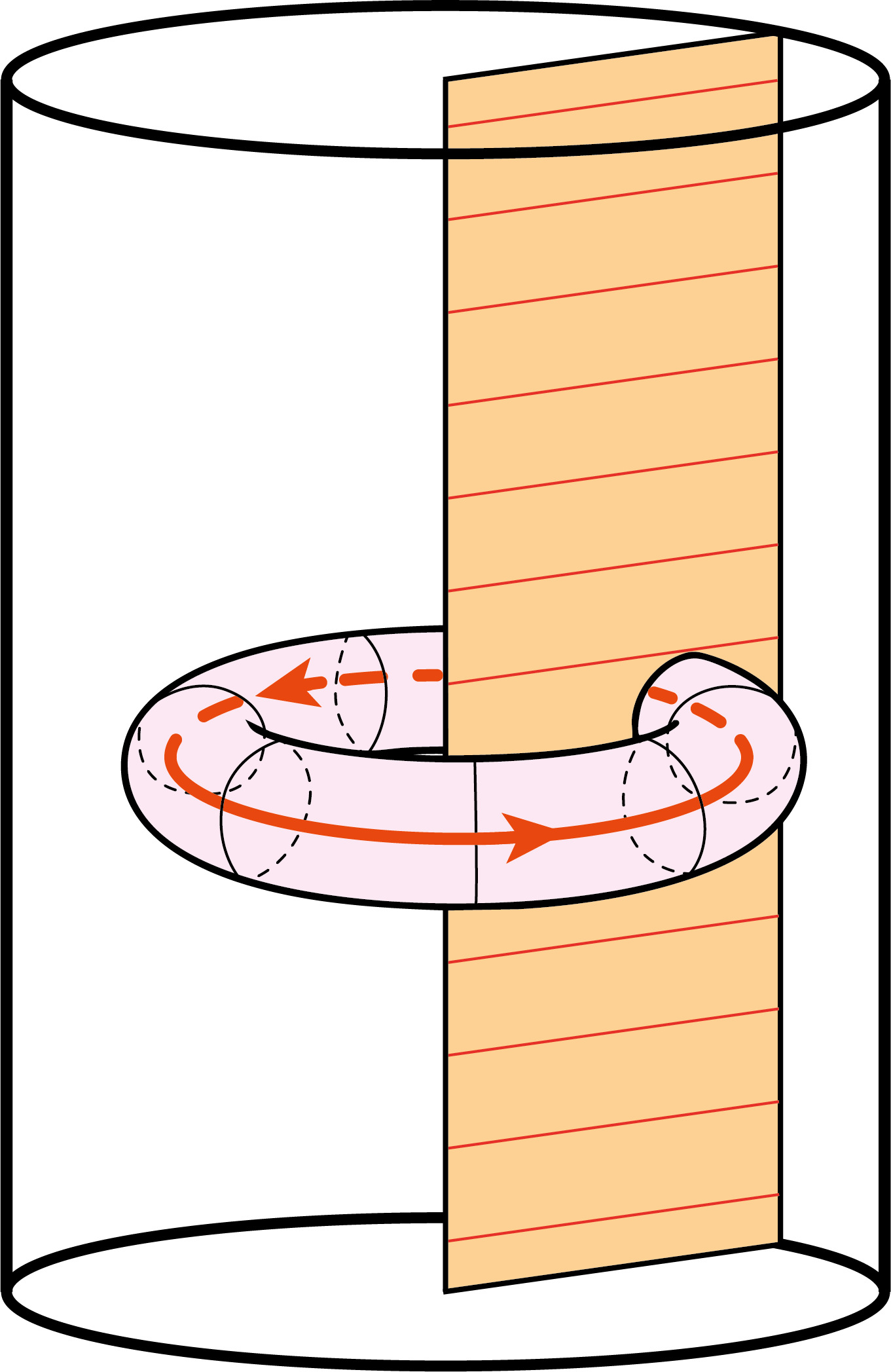}};
    \end{tikzpicture}
    \caption{A toroidal neighborhood of the transverse unknot}
    \label{fig:tor}
\end{figure}

In a small toroidal neighborhood of the knot, the layers of the smectic intersect this neighborhood along meridian disks, as shown in Figure \ref{fig:tor}.

\begin{figure}[!ht]
    \centering
    \begin{tikzpicture}
    \node at (0,0) {\includegraphics[scale=0.25]{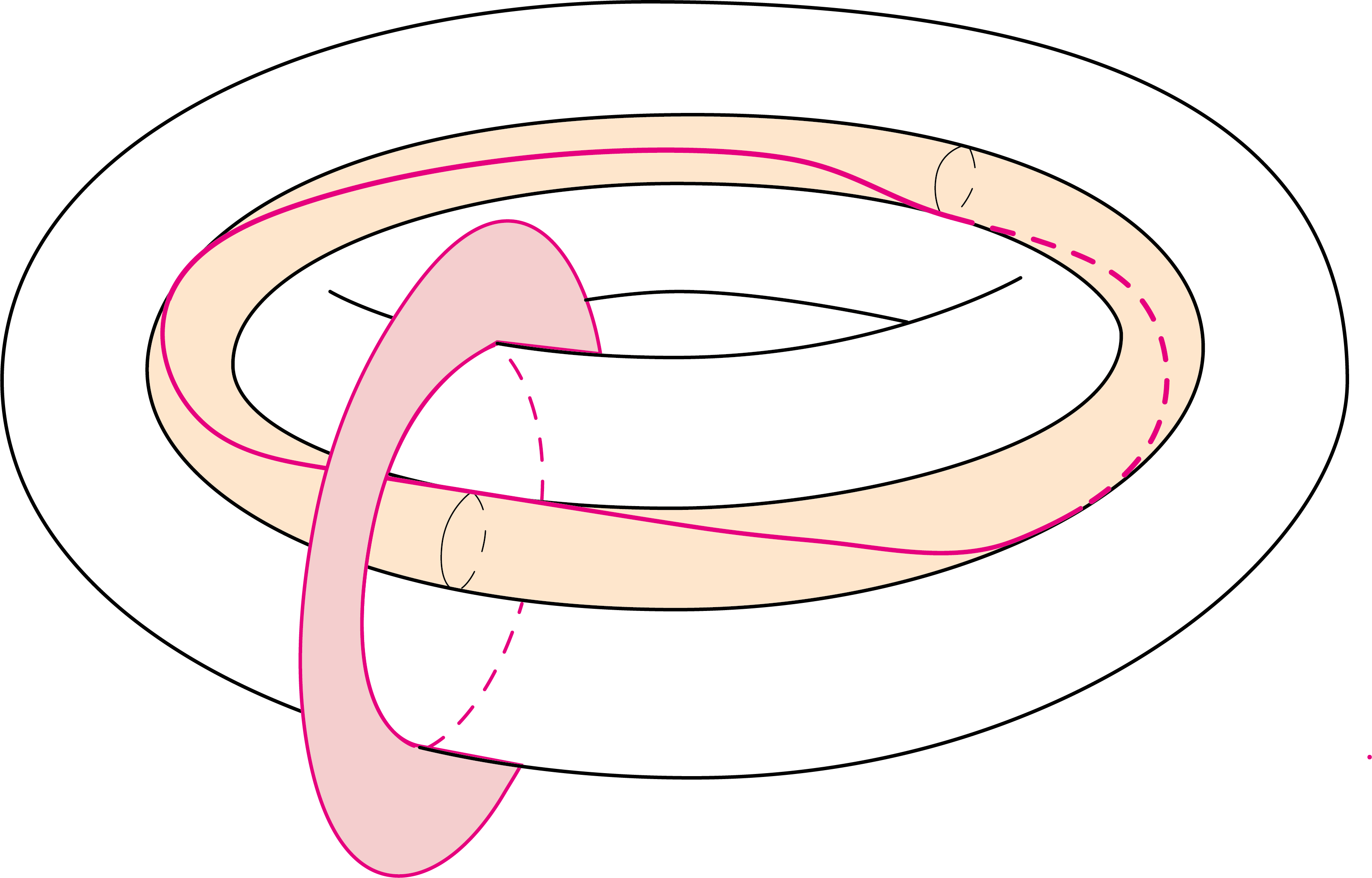}};
    \end{tikzpicture}
    \caption{The toroidal shell and its boundary conditions}
    \label{fig:torus}
\end{figure}

Thus, we can identify this neighborhood of the knot again with a cylinder (with the top and bottom disks identified) where the layers are given by meridian disks. In these coordinates, the trefoil knot corresponds to the core line (or loop) of the cylinder. To transform this core loop into a new dislocation loop of the smectic, we proceed as follows. We consider a slightly thinner toroidal neighborhood of the core, and fill it with a smectic configuration whose layers are given by helicoids (as in the right side of Figure \ref{fig:rational}). The core of this thinner neighborhood, which corresponds to the transverse trefoil knot, is a loop singularity of this local configuration. A ``toroidal shell" lies between the boundaries of the thinner and thicker neighborhoods of the knot, and its boundary is given by two cylinders (or tori after identification) of different radii. The situation is depicted in Figure \ref{fig:torus}. The layers of the ambient smectic configuration, which is fixed away from the thick neighborhood of the trefoil knot, intersect the torus with a large radius along meridian circles, as depicted by the red surface in the figure. On the other hand, the helicoidal layers of the thin neighborhood intersect the torus of small radius along curves that wrap once around the knot, like the one drawn in red in the figure.

\begin{figure}[!ht]
    \centering
    \begin{tikzpicture}
        \node at (0,0) {\includegraphics[scale=0.25]{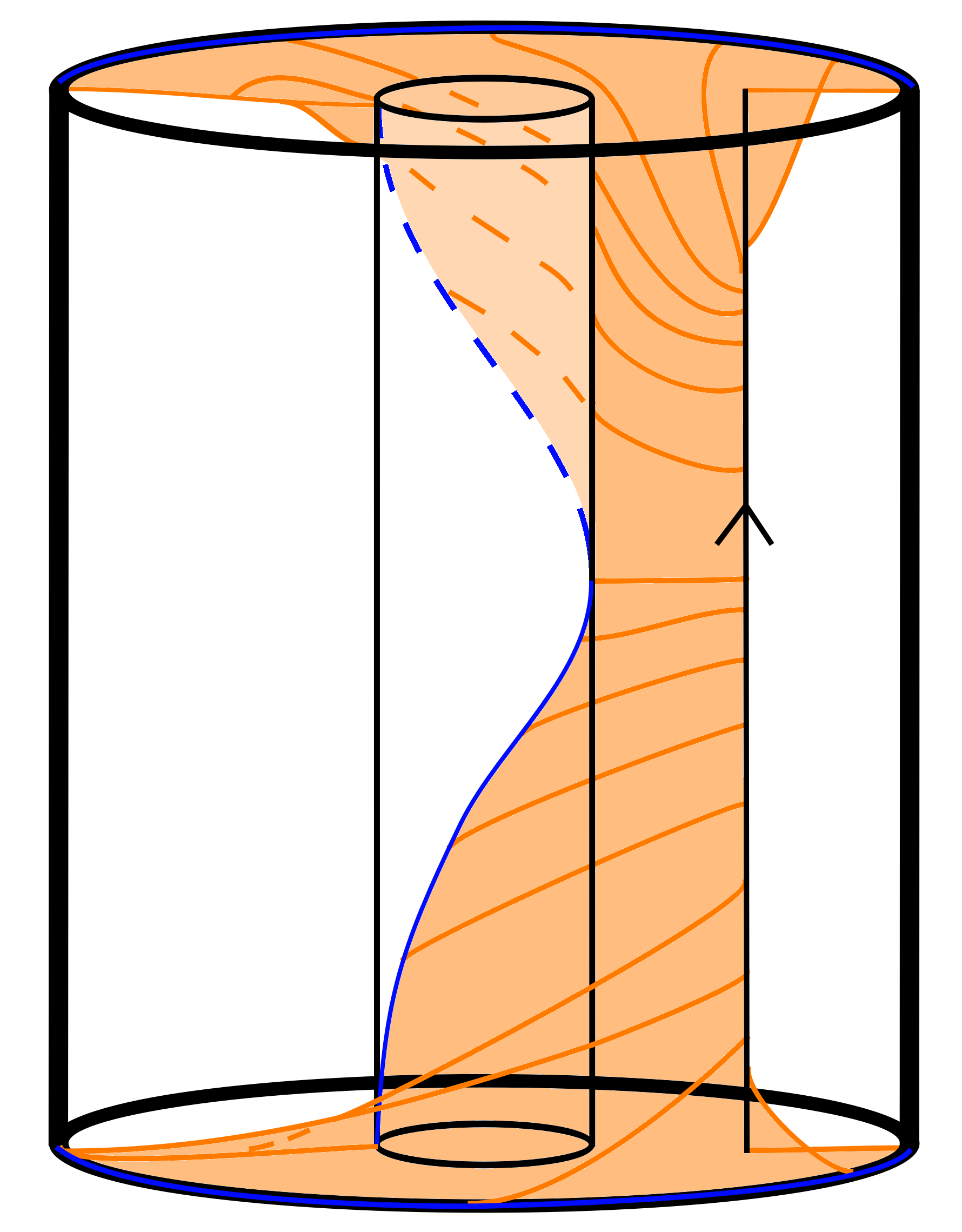}};
    \end{tikzpicture}
    \caption{A surface in the toroidal shell with a new line singularity}
    \label{fig:smooth1}
\end{figure}

\medskip 
We are left with the task of filling this toroidal shell with surfaces that match up with these two boundary conditions. It is impossible to fill it with a smooth family of surfaces unless a new singularity is introduced. We consider ruled surfaces on the toroidal shell like the one depicted in orange in Figure \ref{fig:smooth1}, where the top and bottom of the cylinder have to be identified. 

Notice that, as intended, this surface intersects the outer cylinder along a meridional circle, and the inner cylinder along a circle that wraps once around the core of the cylinder. However, it has another boundary component that hits a closed curve (depicted by a vertical line directed by an arrow) in a helicoidal fashion, introducing a new line singularity in the toroidal shell. This new closed curve is a curve that is isotopic to the core circle and thus is a parallel copy of the transverse trefoil knot. Considering parallel copies of the orange surface, we end up with a smectic configuration that is now defined globally, matches the original one away from a neighborhood of the knot, and that has two new dislocation loops: the transverse trefoil knot and a parallel copy of it. One might check that, as an additional observation, the layers wrap in opposite orientations on these two new loop singularities.

\medskip
With a more intricate construction involving more general helicoidal-like surfaces (used in both the shell and in the neighborhood of the new loop singularity), this very same procedure can be proven to work for any starting transverse knot (or link) in any smectic configuration. By using these surfaces in different manners, one can also prescribe a new topological invariant (the ``local multiplicity") of loop dislocations that we introduce in Section \ref{fig:mult}. This invariant captures, in some sense, how the layers behave near the loop singularity. To picture an example, the multiplicity of the singularities in Figure \ref{fig:rational} is one since the boundary of each layer is mapped once (meaning in a one-to-one manner) to the loop singularity. On the other hand, the regular layers of the configuration in Figure \ref{fig:mult} have a boundary that is mapped in a two-to-one manner to the loop singularity. \\

Lastly, a natural question arises. Given any knot or link, can we find a copy of it (meaning another embedded curve knotted in the same way) that is transverse to the layers of a given smectic configuration? If the answer were positive, by our construction, it would mean that any knot can be introduced to the set of dislocations by a local procedure. The last section of this work shows that for certain smectic configurations, such as those arising from so-called ``open book decompositions" (examples of those are constructed in \cite{SKB}), the answer is indeed positive: given any starting knot (not necessarily transverse to the layers) it is possible to deform it, without changing the way in which it is knotted, so that it becomes transverse to the layers of the smectic. While this construction is harder to sketch informally (as it relies on the technical definition of open book decomposition), let us briefly mention how this is achieved in two steps. To set up ideas, a loose description of a smectic configuration given by an open book decomposition is a configuration for which the regular layers (in the complement of the loop singularities) are obtained by taking parallel copies of a surface along a circle direction that winds around the singularities. This is the case for the two configurations on the cylinder depicted in Figure \ref{fig:rational} (where the circle direction would go around the $z$ axis). In the first step of the construction, we slightly move the knot so that it is in ``general position" with respect to the layers, which means that it does not intersect the loop singularities, and that it is tangent to the regular layers only at finitely many points. Having chosen an orientation of the layers, this leaves us with a knot that is somewhere positively transverse to the layers, somewhere transverse in the opposite direction to the layers, and tangent at some points. The second step of this proof involves deforming the knot iteratively on each part that is transverse in the "wrong" direction, and making it positively transverse while undoing the tangency points. \\

All put together, our local construction of loop dislocations is shown to be rich both in a global topological sense (the manner in which the loop is knotted can be arbitrary) and a local sense (the layers can wrap around the knot arbitrarily).

\section{Smectics and open books} \label{s:back}
In this section, we recall the mathematical objects that model smectic liquid crystals, their defects, and the notion of rational open book decomposition of a three-manifold.

\subsection{Smectic liquid crystals and screw dislocations}
The mathematical model for an ordered medium in a configuration space $M$ is given by an order parameter
$$\Phi: M \longrightarrow V$$
that characterizes the internal state of the medium at any point \cite{TK}. In the case of 3D liquid crystals, the configuration space is a three-dimensional manifold $M$ endowed with some ambient Riemannian metric, and the order parameter is a line field $\mathbf{n}$ in $M$, a one-dimensional subbundle of $TM$, known as the \emph{director line field}. The map $\Phi$ might be defined in the complement of a set of points in $M$ known as the defects (on the director) of the medium, which are either isolated points or lines. \\

Smectic liquid crystals in three dimensions admit a layered structure, defined everywhere except along a set of singularities. The molecules are bound to move along these layers and are oriented in the direction orthogonal to the layers. This layered structure is modeled by a singular foliation $\mathcal{F}$ for which $\mathbf{n}$ is orthogonal to the regular codimension one leaves. The singularities of the foliation can be either points or lines, but line defects seem to play a more structural role \cite{MAHK}. Among the types of defects of a smectic, dislocations are those where the foliation is singular without a discontinuity on the director field. Two types of line dislocations are screw and edge dislocations, which can be distinguished topologically \cite{KM}. A screw dislocation is a line (or a loop) for which each layer near any point in the line takes the form of a helicoid.

Since the regular leaves of $\mathcal{F}$ model the evenly-spaced layers of a smectic, we will always work with foliations for which the closure of each regular leaf is compact when the ambient space $M$ is compact. In $\mathbb{R}^3$, one can require instead that in the compact set $K\subset \mathbb{R}^3$ where $\mathcal{F}$ does not necessarily match the ground state, the closure of each leaf is compact. This is always the case for instance in \cite{SKB}.

\subsection{Rational open book decompositions} In Section \ref{s:link} we will consider smectic configurations all of whose singularities are given by screw dislocations, formed along knots. The layered structure $\mathcal{F}$ on a compact three-manifold $M$ (such as $S^3$ or the three-torus $T^3$) will be given by a rational open book decomposition. From a physical point of view, one can keep in mind the case of a rational open book decomposition of $S^3$, which, up to projecting as in \cite{SKB}, gives a smectic configuration in $\mathbb{R}^3$ matching the ground state away from a ball. Let us then introduce a bit of notation and the formal definition of a rational open book decomposition. 

Given a knot $K$ on a three-manifold $M$, we can choose a tubular neighborhood of the form $U\cong S^1\times D^2_\delta$, where $D^2_\delta$ is a closed disk of radius $\delta>0$. A meridian of this neighborhood is given by the boundary of one of the disks $\{p\}\times D^2$, while a longitude (or framing curve) is a curve in $\partial U$ isotopic to the knot $S^1\times \{0\}$. Up to choosing well the coordinates of this neighborhood, one can assume that the longitude is given by $S^1\times \{q\}$, for some point $q$ in the boundary of $U$.

\begin{defi}
Let $M$ be a compact manifold (possibly with boundary). A \emph{rational open book decomposition} $(B, \pi)$ of $M$ is a link $B$ (not intersecting the boundary of $M$) and a fibration
$$\pi: (M\setminus B) \longrightarrow S^1,$$
transverse to the boundary of $M$ and satisfying the following property. In the neighborhood $U$ of any connected component $K\subset B$, each fiber $f^{-1}(\theta)$ intersects $\partial U$ along a curve that is not a meridian. 
\end{defi}

Thus, each fiber of $\pi$ intersects $\partial U$ along an embedded collection of closed curves with a fixed singular homology class in $H_1(\partial U;\mathbb{Z})\cong H_1(T^2;\mathbb{Z})\cong \mathbb{Z}^2$. This group is generated by the homology class of a meridian and a longitude. The class of the collection of curves induced by a fiber always has a non-vanishing longitudinal component, and it is primitive if and only if this collection has a single element. In the rest of this article, each time we speak of an open book decomposition, we mean a rational open book decomposition.

If we identify a neighborhood of the binding with a solid torus formed by gluing the two ends of the cylinder together, Figure \ref{fig:rational} pictures two possible ways in which the pages can behave near the binding. The left side corresponds to the binding of a standard open book decomposition, and the homology of each page in the boundary is that of a longitude. On the right side is an example of how the pages can behave around the binding in a rational open book decomposition, in this case, the homology of each page in the boundary is that of the sum of a longitude and a meridian. Many more pictures as possible, namely, a page can wrap arbitrarily many times around the binding or in the binding direction. We will come back to this fact when discussing the local multiplicity in Section \ref{ss:mult}.

 \section{Screw dislocations near a transverse link}\label{s:local}

In this section, we describe a geometric construction to locally introduce screw-like knotted defects in a given smectic along a transverse knot.

\subsection{The setup}
Suppose that we are given a smectic configuration with singular foliation $\mathcal{F}$. We fix a knot (or a link, the construction is analogous) $K$ that is assumed to be transverse to the regular leaves of the foliation. Notice that this can never be the case in the ground state in $\mathbb{R}^3$, since all the layers are horizontal planes, and such a foliation does not admit an embedded transverse circle. Since $K$ is transverse to the foliation, locally near any point of $K$ we can take a foliated chart where the leaves are given by small disks transverse to the knot. Thus, if we consider a small enough toroidal neighborhood $U\cong D^2\times S^1$ of $K$, any leaf of $\mathcal{F}$ necessarily intersects $U$ along a collection of disks transverse to $S^1$ with boundary in $\partial U$. We aim to modify the layers in $U$ by introducing $K$ as a new screw-like line defect of the smectic.

\subsection{Helix boxes}\label{ss:helix}
\begin{figure}[!ht]
    \centering
    \begin{tikzpicture}
    \node at (0,0) {\includegraphics[scale=0.09]{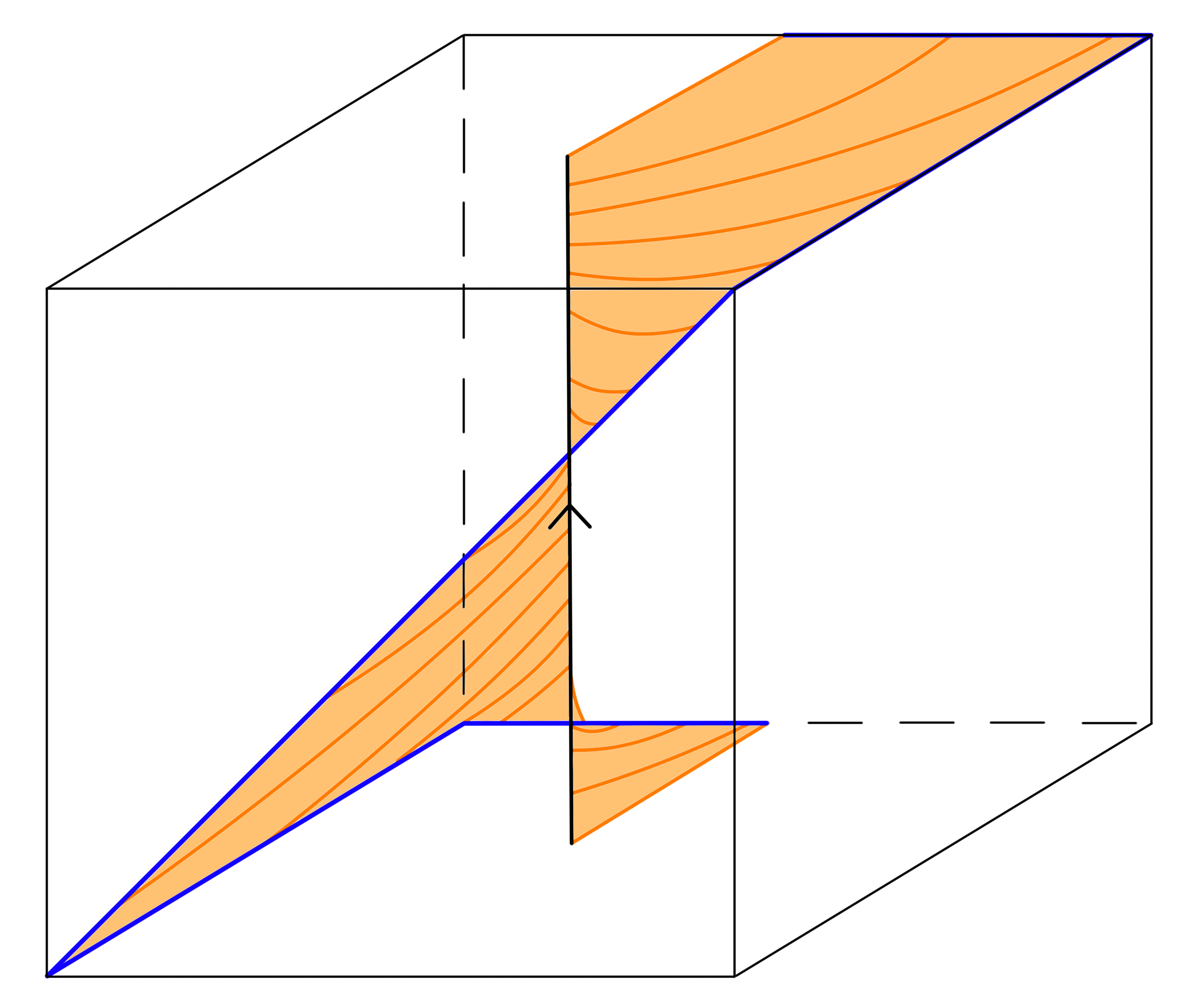}};
    \node at (6.5,0) {\includegraphics[scale=0.09]{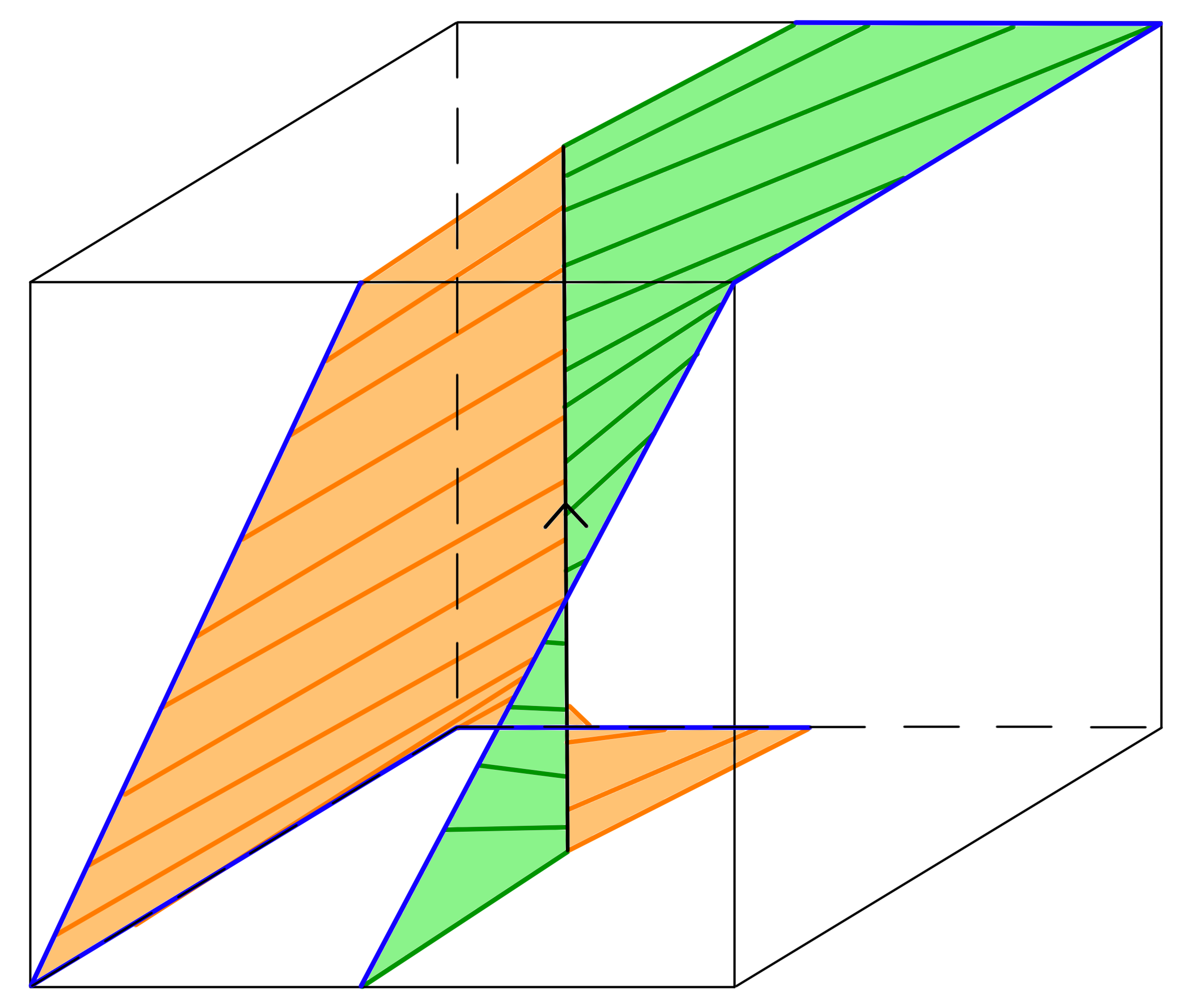}};


    \draw[->] (2.4+0.2,-2.2) -- (1.8+0.2,-2.2);
    \node[scale=0.8] at (1.9+0.2, -2.5) {$x$};

    \draw[->] (2.4+0.2,-2.2) -- (2.4+0.2, -1.6);
    \node[scale=0.8] at (2.65+0.2, -1.55) {$y$};

    \draw[->] (2.4+0.2,-2.2) -- (2.85+0.2, -1.85);
    \node[scale=0.8] at (2.9+0.2, -2.15) {$t$};
    \end{tikzpicture}
    \caption{Helix boxes}
    \label{fig:helix}
\end{figure}

Before proceeding with modifying the layers, we recall first the main tool that we will use: ``helix boxes". We follow the definition in \cite{CR}, where the most general version of helix boxes is first introduced (the first appearance of a simpler version goes back to \cite{VH}). A helix box $(N,\Sigma)$ is a domain $N$ diffeomorphic to $T^2\times I$, where $I=[0,1]$, and a certain immersed surface with boundary $\Sigma$ on $N$. The surface is embedded except perhaps along some of the boundary components that are mapped to a closed embedded curve in the interior of $N$. To specify the surface in the helix box, we need to fix the following data:
\begin{itemize}
    \item[-] a finite embedded collection $\sigma_1$ of closed curves lying in $T^2\times \{1\}$ and inducing a non-trivial class $[\sigma_1]$ in $H_1(T^2\times I; \mathbb{Z})$,
    \item[-] a closed embedded curve $\gamma$ lying in $T^2\times \{1/2\}$, non-trivial in $H_1(T^2\times I; \mathbb{Z})$ and linearly independent of $[\sigma_1]$,
    \item[-] a finite embedded collection $\sigma_0$ of closed curves lying in $T^2\times \{0\}$ such that $[\sigma_0]=[\sigma_1]+ p [\gamma]$ for some integer $p\neq 0$.
\end{itemize}
 Given this data, we can construct the surface $\Sigma$ as follows. Write $[\sigma_1]=q[\sigma]$ where $\sigma$ is a connected embedded curve. Then $[\sigma],[\gamma]$ determine a basis of $H_1(T^2\times I; \mathbb{Z})$. Identify $T^2\times I$ with a cube $I^3$ with coordinates $(x,y,t)$ and the equivalence relation: 
\begin{equation*}
    \begin{cases}
        (x,1,t) \sim (x,0,t), \enspace x,t\in [0,1],\\
        (1,y,t) \sim (0,y,t), \enspace y,t\in [0,1].
    \end{cases}
\end{equation*}
Choosing well the identification of $T^2\times I$ with $I^3/\sim$, we can assume that each curve in $\sigma_1$ is parallel to $\pp{}{x}$ and that $\gamma$ is parallel to $\pp{}{y}$ (for instance, it corresponds to $(1/2, y, 1/2)$ as in Figure \ref{fig:helix}). This is possible because $\sigma$ and $\gamma$ define a basis of $H_1(T^2\times I;\mathbb{Z})$.

The curves $\sigma_1$ and $\sigma_0$ give rise to a collection of piecewise linear segments in $I^2\times \{1\}$ and $I^2\times \{0\}$ respectively. We add straight segments in $\{0\}\times I^2$ and $\{1\}\times I^2$ every time there is a point of $\sigma_1$ and of $\sigma_0$ in one of these two faces of the cube at the same height (with same $y$ coordinate). We obtain in this way a collection of piece-wise linear curves like those drawn in blue in Figure \ref{fig:helix}. In the left side example, we have $q=1$ and $p=-1$: the curve $\sigma_1$ in $I^2\times \{1\}$ is horizontal (we have drawn half of this segment on each side of the identified faces), the curve $\sigma_0$ is diagonal, and we add a segment in $\{0\}\times I^2$ and one in $\{1\}\times I^2$. The right side example corresponds to $q=1$ and $p=-2$. If $q=2$ and $p=-1$, there would be one piece-wise linear curve that would wrap twice around the box horizontally, see a figure in \cite[p 265]{CR}.

Each of these segments can then be used to span ruled surfaces connecting the central curve $\gamma$ with each one of the piecewise segments in the boundary of $I^3$. Concretely, choose polar coordinates $(r,\theta)$ in the $(x,t)$ plane. We parametrize one of the piecewise linear segments by $s:[0,1] \rightarrow I^3$, where $s(v)=(r(v), y(v), \theta(v))$. Then, a piecewise version of the surface is given by a surface that in coordinates $(r,y,\theta)$ takes the form
$$\varphi(u,v)= \big(ur(v), v , \theta(v)\big), \quad u\in [0,1], v\in [0,1].$$
One might easily smooth out this surface at convenience. By construction, all these surfaces together give a well-defined (possibly non-connected) immersed surface $\Sigma$ with boundary in the quotient $T^2\times I$ with the following properties:
\begin{itemize}
    \item[-] $\partial \Sigma\cap T^2\times \{1\}=\sigma_1$,
    \item[-] $\partial \Sigma \cap T^2\times \{0\}=\sigma_0$,
    \item[-] $\Sigma$ is embedded except possibly along $\gamma$, where a subset of the connected components of the boundary of $\partial \Sigma$ are mapped via an immersion.
\end{itemize}
Notice that parallel copies of $\Sigma$ in the $y$ direction define a fibration of $\left(T^2\times I \right)\setminus \gamma$ over the circle. It is then the page of rational open book decomposition of $T^2\times I$ with binding $\gamma$.

\subsection{Local modification of layers}\label{ss:usehelix}

To set up ideas, we will first show how to use helix boxes to introduce knotted screw loop defects. We will later modify this construction to prescribe some topological invariants along the screw loop defect.\\

Recall that in a neighborhood $U\cong D^2\times S^1$ of $K$, each leaf of the foliation $\mathcal{F}$ intersects $U$ along finitely many disks with boundary on $\partial U$. In particular, we can choose different coordinates $(x,y,\theta)$ in $U\cong D^2\times S^1$ so that each leaf of the foliation $\mathcal{F}$ intersects $U$ along a finite union of disk fibers of the form $\theta=\{c\}$. Notice that there is no way of introducing a single screw dislocation along $K$ in a way that the new smectic configuration in $U$ matches this boundary condition. To resolve this issue, we will introduce two screw dislocations, one along $K$, and one along a parallel copy of $K$.\\

\begin{figure}[!ht]
    \centering
    \begin{tikzpicture}
        \node at (0,0) {\includegraphics[scale=0.35]{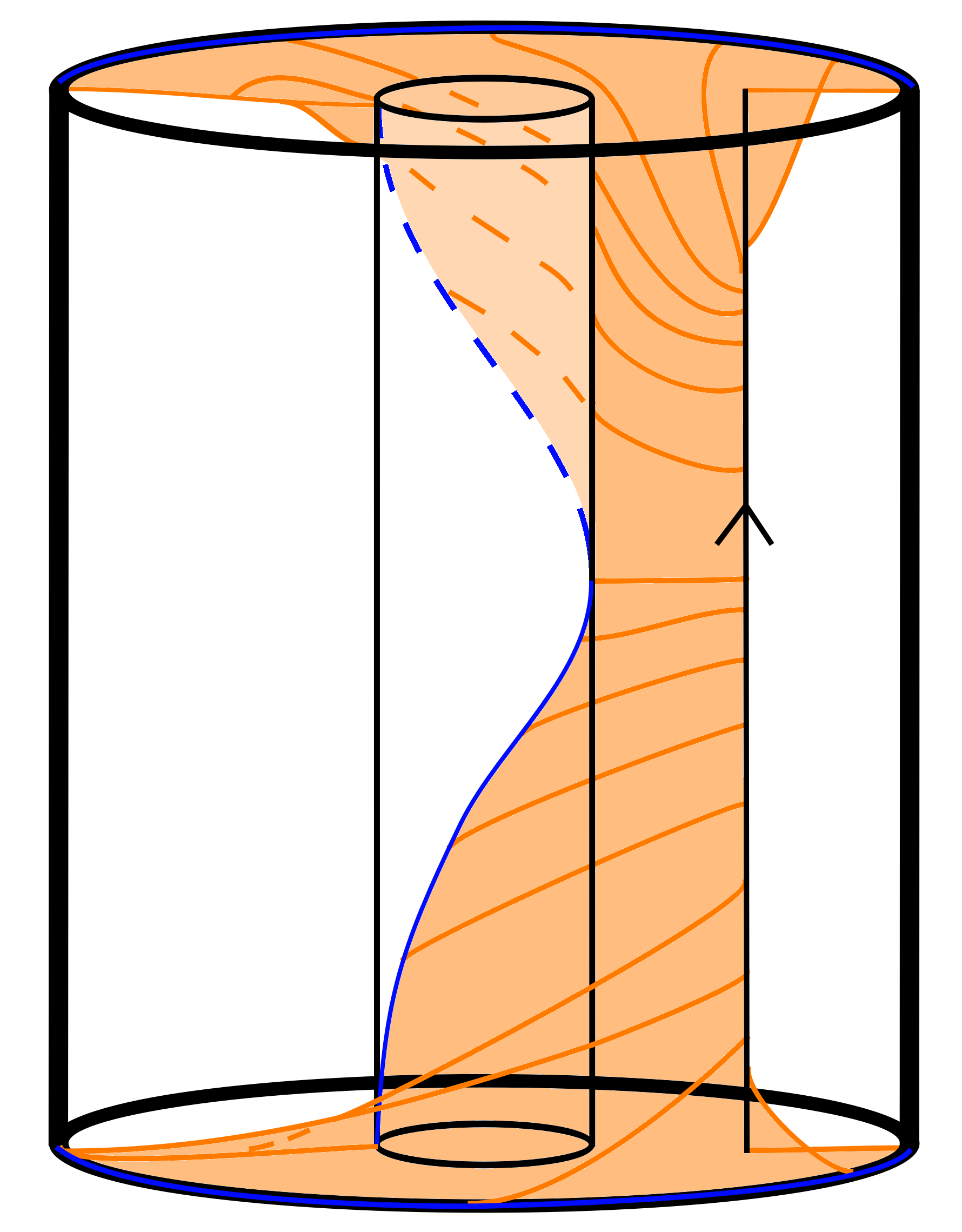}};
        \draw[|->] (0.65,-3.5)--(2.3,-3.5);
        \node at (1.45, -3.8) {$t$};
    \end{tikzpicture}
    \caption{Smooth helix box in $V$}
    \label{fig:smooth}
\end{figure}

Take a curve $\sigma_1$ to be the boundary of one of the disk fibers, and a curve $\gamma$ to be $\{p\}\times S^1$ with $p$ of radius $1/2$. The curve $\gamma$ is then a curve parallel to the core of $U$. Choose some small enough $\delta>0$, and consider the set
$$V=U\setminus (D^2_{\delta}\times S^1) \cong [\delta,1]\times T^2.$$
In this domain, the pair $[\gamma], [\sigma_1]\in H_1(V;\mathbb{Z})$ defines a basis of the first homology group of $V$, with $\gamma\subset T^2\times \{1/2\}$ and $\sigma_1\subset T^2\times \{1\}$. By construction, each leaf $L$ of the foliation $\mathcal{F}$ intersects $\partial V$ along $k$ closed curves, and each one is the boundary of one of the disk fibers of $V$. In particular, each leaf satisfies $[L\cap \partial V]=k[\sigma_1]$. We will now modify the leaves of $\mathcal{F}$ first in $V$, and then extend this modification inside $U\setminus V$ in a way that the core of $U$ (which is $K$) will become a screw-like loop defect of the new smectic configuration. To do so, we consider a linear curve $\sigma_0 \in T^2\times \{\delta\}$ such that $$[\sigma_0]=[\sigma_1]+[\gamma].$$
Parallel copies of the surface $\Sigma$ of a helix box with data $\sigma_0, \sigma_1, \gamma$ can then be used to extend the layers of the smectic in $M\setminus U$ inside $V$. See Figure \ref{fig:smooth} for a smooth version of one of these surfaces, where we think of $T^2\times I$ as $S^1\times I \times I$ and we identify the top and bottom annuli. 

We obtain a smectic configuration defined in $N=M\setminus \left(D^2_\delta\times S^1\right)$, and if the original smectic is a rational open book, so is the smectic configuration in $M\setminus \left(D^2_\delta \times S^1\right)$. In $U'$, the closure of $D^2_\delta \times S^1$, a layer of the smectic in $N$ intersects $\partial U'$ along $\sigma_0$. The curve $\sigma_0$ is not meridian, since $[\sigma_0]$ has a non-vanishing longitudinal component. In particular, one can extend the surface inside $U'$ as a helicoid $H$ having the core of $U'$ as a spanning curve, thus filling the cylindrical (which, after identification, is toroidal) hole in Figure \ref{fig:smooth}. Using parallel copies of the helicoid, we extend every layer inside $U'$. We obtain a new smectic configuration, i.e., a singular foliation $\mathcal{F}'$ in $M$, which coincides with $\mathcal{F}$ away from $U$. This new foliation has two new loop singularities, namely, the knot $K$ and the curve $\sigma \subset T^2\times \{1/2\}$. Near these singularities, the layers wind around in a helicoidal fashion. This finishes the construction of a new smectic having $K$ as a screw loop defect. 

\begin{Remark}
    Notice that if we orient $K$ and $\gamma$ consistently (in a way that one is a translation of the other as oriented curves), the helicoid near $K$ twists in one direction and near $\gamma$ in the opposite one. 
\end{Remark}

\begin{Remark}
    The argument in this section has as a consequence, in terms of knot theory, that if $B$ is a fibered link and $L$ is a link transverse to the fibers of the associated fibration, then $B\cup L \cup L'$ is fibered. Here $L'$ is a translated copy of $L$ with the opposite orientation. A version of this statement can be found in \cite[Theorem 2]{HMS} with an algebraic rather than geometric proof.
\end{Remark}

\section{Prescribing the multiplicity invariants}\label{s:localmult}
In this section, we introduce a topological invariant of a screw dislocation and adapt the construction of Section \ref{s:local} to be able to prescribe it.

\subsection{Multiplicity along screw loop defects}\label{ss:mult}
\begin{figure}[!ht]
\centering
    \begin{tikzpicture}
        \node at (0,0) {\includegraphics[scale=0.35]{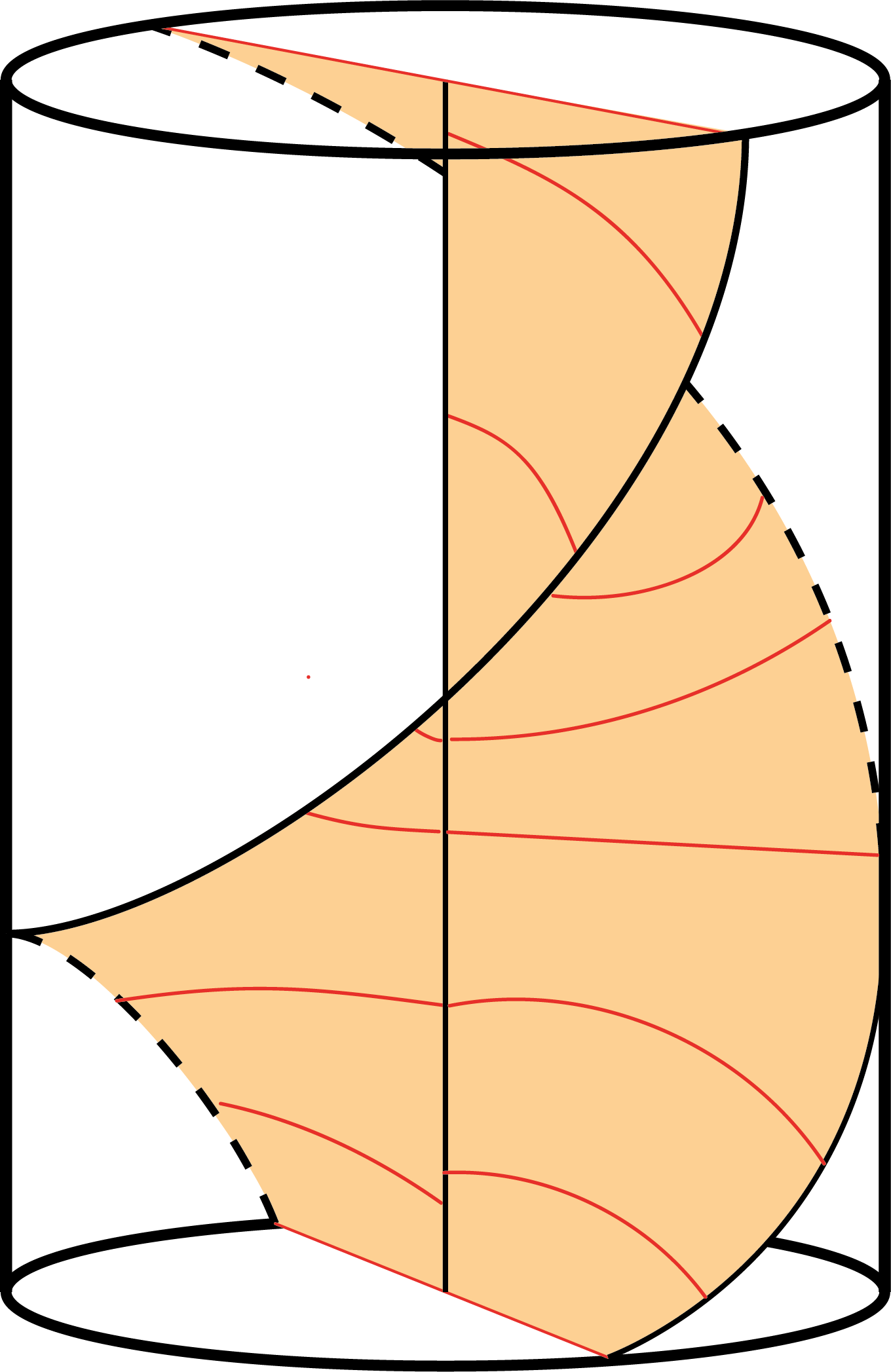}};
        \node at (6,0) {\includegraphics[scale=0.35]{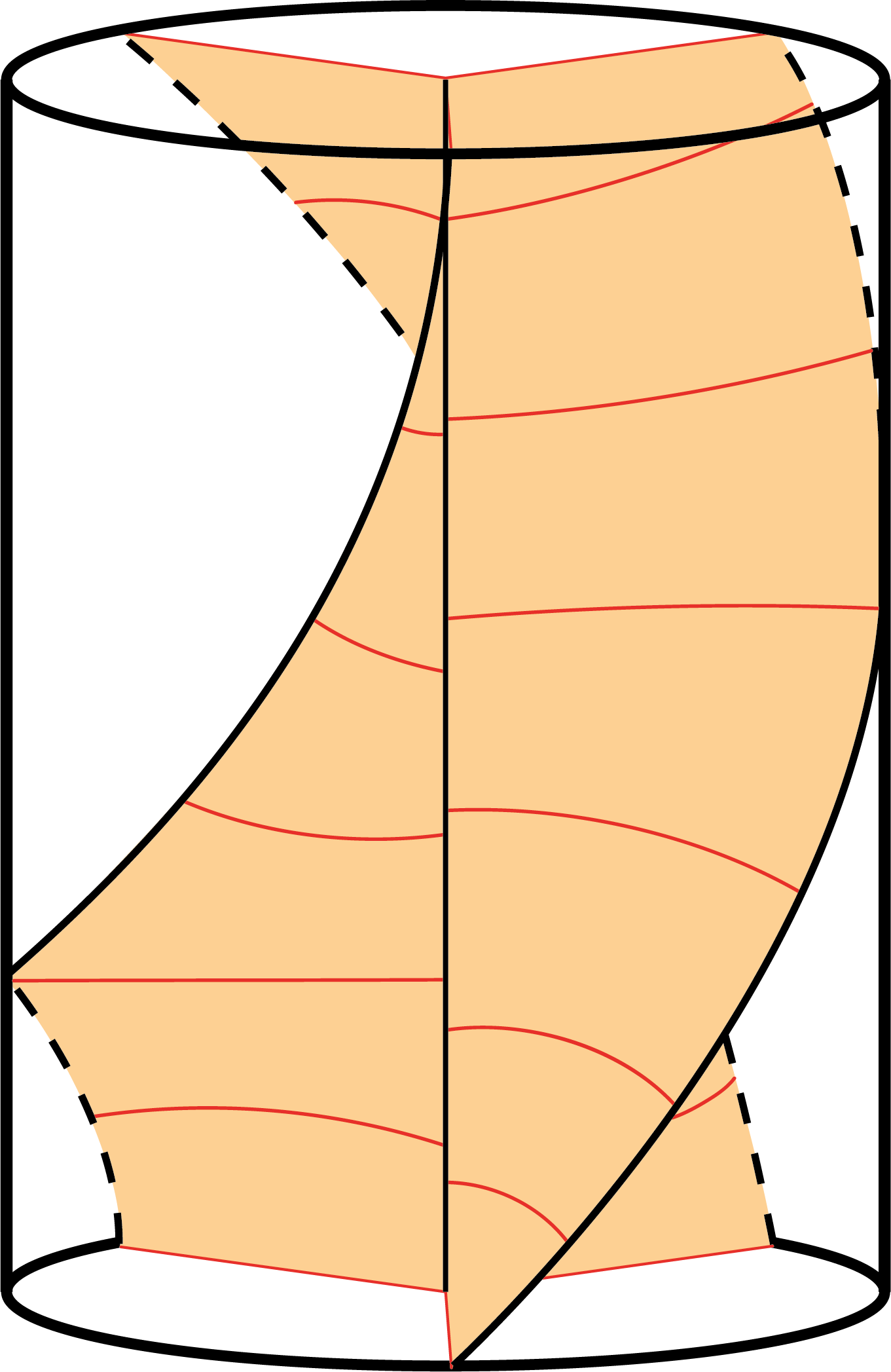}};
    \end{tikzpicture}
    \caption{Bindings with multiplicity $2$ and $3$}
    \label{fig:mult}
\end{figure}

Let $B$ be a screw loop defect of a smectic with foliation $\mathcal{F}$. There is a neighborhood $U\cong D^2\times S^1$ of $B$ where $\mathcal{F}$ is a fibration over $S^1$ in $U\setminus B$ (whether $\mathcal{F}$ is given by an open book or not). For simplicity, we choose the fibration so that each fiber is connected. A fiber is then always an immersed annulus
$$\iota_A: A \longrightarrow U,$$
which is an embedding except maybe along the boundary component $\nu\subset \partial A$ mapped to $B$, the other boundary component being mapped to $\partial U$. From now on, let us assume that the smectic director line field is orientable, in which case each layer admits a natural orientation induced by the orientation of the director field, and $B$ gets an induced orientation too. The restriction of the map $\iota_A$ to $\nu$ is then a finite covering map of a circle to $B$. Fixing the orientations, the degree of the restriction of $\iota_A$ to $\nu$ is an integer-valued topological invariant of the singularity. 
\begin{defi}
    The \emph{local multiplicity} of $\mathcal{F}$ along $B$ is $\operatorname{m}_{\mathcal{F}}(B):=\operatorname{deg}(f)$, where 
    $$f:=\iota_A|_{\nu}: S^1 \longrightarrow B\cong S^1,$$
    for any choice of connected regular layer $A$ of $\mathcal{F}$ in any sufficiently small neighborhood $U$ of $B$. If the director field is not cooriented, we define it instead as $\operatorname{m}_{\mathcal{F}}(B):=|\operatorname{deg}(f)|$.
\end{defi}
Notice that this is well-defined, i.e., does not depend on the choice of the layer, since the layers define a fibration over $S^1$ in the complement of $B$ in $U$, and thus $\operatorname{deg}(\iota_A|_{\nu})$ is the same for any fiber $A$. Both examples in Figure \ref{fig:rational} have multiplicity one. On the other hand, Figure \ref{fig:mult} represents a neighborhood of a binding with multiplicity $2$ and $3$ respectively. Notice how the boundary of a single layer covers twice the binding, since the top and bottom disks are identified by the identity map. We point out as well that the sign of the multiplicity determines the direction in which the helicoidal surfaces wind around $B$. 

One can also define the (``global") multiplicity of a regular layer $F$ of the foliation $\mathcal{F}$ along $B$. By this, we mean that a layer $F$ defined in all the ambient space might intersect a small enough neighborhood $U$ of the singularity along a finite collection of fibers $A_1\sqcup ... \sqcup A_r$ of the annulus fibration structure in $U\setminus B$. For example, on the right side of Figure \ref{fig:rational}, locally the orange and blue surfaces are different fibers of the annulus fibration, while they could correspond to the same layer globally. If each fiber of the fibration has a boundary component that covers $\operatorname{m}_{\mathcal{F}}(B)$ times the knot $B$, the boundary components of the compactification of $F$ that are mapped to $B$ cover it $r\operatorname{m}_{\mathcal{F}}(B)$ times (that is, $m$ times for each fiber that corresponds to the same layer).
\begin{defi}
    The \emph{(global) multiplicity} of a layer $F$ along $B$ is $\operatorname{m}_F(B):=r\operatorname{m}_{\mathcal{F}}(B)$, where $r$ is the number of connected components of $F\cap U$ for any sufficiently small neighborhood $U$ of $B$.
\end{defi}
In the next section, we will see a construction that produces plenty of examples of smectics for which the global multiplicity of a layer along a screw loop defect does not match the local multiplicity. In general, the global multiplicity along a screw defect might have different values for different layers if $\mathcal{F}$ is not given by a fibration over $S^1$ in the complement of the singular leaves.

\subsection{Helix boxes to prescribe multiplicities} \label{ss:presc}
Let us discuss how the construction in Section \ref{ss:usehelix} can be modified to prescribe the values of the invariants that we just introduced. 

\paragraph{Prescribing the local multiplicity.} Recall that, as in Section \ref{ss:usehelix}, we choose a knot $K$ that is transverse to the layers of a smectic configuration given by a foliation $\mathcal{F}$. There is a small enough neighborhood $U\cong D^2\times S^1$ of $K$, such that the connected components of the layers along $U$ are of the form $D^2\times \{p\}$ with $p\in S^1$. Choose a small enough $\delta>0$, and equip $\overline{D^2_\delta} \times S^1$ with a helicoidal smectic configuration $\mathcal{F}_k$ with multiplicity $k$, i.e. the regular leaves are helicoids that cover the screw loop defect $\{(0,0)\}\times S^1$ exactly $k$ times. Such examples can be pictured as generalizations of Figure \ref{fig:mult}. In that example, the helicoid rotates an angle of $\pi$ each time it turns once in the direction of $K$. To obtain a helicoid of multiplicity $k$, one simply has to consider layers that turn an angle of $\frac{2\pi}{k}$ as they go turn once in the direction of $K$.

As we did in Section \ref{ss:usehelix}, we can choose two curves $\sigma_1, \gamma$ in the domain
$$V=U\setminus \left(D^2_\delta \times S^1\right)\cong T^2\times [\delta,1]$$
that generate $H_1(V;\mathbb{Z})$, where $\sigma_1$ is the intersection of a layer of $\mathcal{F}$ with $T^2\times \{1\}$, and $\gamma \subset T^2\times \{1/2\}$ is parallel to the knot $K$. A layer of $\mathcal{F}_k$ intersects $T^2\times \{\delta\}$ along a curve $\sigma_0$. Without loss of generality, we can choose $\mathcal{F}_k$ so that
$$[\sigma_0]= [\sigma_1] + k [\gamma],$$
by choosing the helicoids in $\mathcal{F}_k$ as we described above, namely, so that the normal vector of the helicoid near $K$ rotates by an angle of $\frac{2\pi}{k}$ each time it goes around $K$. We can then use a helix box in $V$ with data $\sigma_0, \sigma_1, \gamma$, and this provides a way of extending the foliation $\mathcal{F}$ inside $V$ so that it matches $\mathcal{F}_k$ in $T^2\times \{\delta\}$. This gives a smectic configuration $\mathcal{F}'$ that coincides with $\mathcal{F}$ away from $U$, and that has two new screw loop defects: one along $K$ of multiplicity $k$, and one along $\gamma$ (that will have multiplicity $-k$). An example of the helix box that is used for $k=2$ corresponds to the right side of Figure \ref{fig:helix}.


\paragraph{Prescribing both multiplicities.} Let us argue how to use the construction to prescribe the local multiplicity along the knot $K$ and the global multiplicity of $F$ simultaneously to be any positive multiple of the local multiplicity. As opposed to the previous constructions, this requires the use of two helix boxes and thus the local formation of three new screw defects, one of them being the transverse knot $K$. \\

From now on, we will use the notation of the previous section. Suppose that we want the local multiplicity to be equal to $k$, and the global multiplicity of $F$ to be $n=rk$. In the neighborhood $U\cong D^2\times S^1$, the layer $F$ intersects the boundary of $U$ along a multi-curve $\Gamma_1$ with $r$ connected components (each of them a boundary of a disk $D^2\times \{c\}$). As before, we equip $\overline{D^2_\delta}\times S^1$ with a helicoidal smectic configuration $\mathcal{F}_k$ of local multiplicity $k$. In $V$, we consider the curves $\sigma_1, \gamma$ (a parallel copy of the knot), which are a basis in $H_1(V;\mathbb{Z})$, and we have $[\sigma_0]=[\sigma_1]+k[\gamma]$. Let $\Gamma_0$ be a collection of $r$ curves parallel to $\sigma_0$ inside $T^2\times \{\delta\}$. Our goal is to find a foliation $\mathcal{F}_V$, possibly with new screw defects in the interior of $V$, which intersects $T^2\times \{\delta\}$ and $T^2\times \{1\}$ along curves parallel to $\sigma_0$ and $\sigma_1$ respectively, and satisfying the following property. There is a connected leaf $F_V$ of $\mathcal{F}_V$ such that 
$$F_V\cap (T^2\times \{1\})= \Gamma_1 \quad \text{and} \quad F_V\cap (T^2\times \{\delta\})=\Gamma_0.$$
If we prove the existence of such a foliation, the foliation $\mathcal{F}$ in $M\setminus U$ extends as $\mathcal{F}_V$ inside $V$ and then as $\mathcal{F}_k$ inside $U\setminus V$. Notice that the resulting smectic, with foliation $\mathcal{F}'$, satisfies that the local multiplicity along $K$ is $k$, and the leaf $F_V$ will have global multiplicity $rk$.
\medskip

Let us then prove the existence of $\mathcal{F}_V$, which will be obtained by using the concatenation of two helix boxes in $V$,  a key idea developed in \cite[Proposition 5.7]{CR}. Take a curve $\gamma' \in T^2\times \{1/8\}$ such that $[\gamma], [\gamma']$ generate $H_1(V;\mathbb{Z})$. Then we can write
$$[\Gamma_0]= [\Gamma_1] + p [\gamma] + q[\gamma'], \quad p,q\in \mathbb{Z}.$$
We will assume that $p$ and $q$ are different from zero, the other cases being much easier (and requiring only one helix box). The idea is to use, first, a helix box in $T^2\times [1/4,1]$ to obtain a surface that intersects the right boundary along $\Gamma_1$ and the left boundary along a curve $\Gamma$ satisfying $[\Gamma]=[\Gamma_0]+q[\gamma']$. This is indeed possible: choose $\Gamma$ to be a linear curve in $T^2\times \{1/4\}$ with such a homology class, and then $\Gamma, \Gamma_1, \gamma$ satisfy the homological conditions listed in Section \ref{ss:helix} to construct a helix box.

Finally, we use a second helix box in $T^2\times [\delta, 1/4]$. This time, we use as data $\Gamma_0, \Gamma$, and $\gamma'$. Notice that $[\Gamma_0]=[\Gamma]-q[\gamma']$, and thus one can again construct a helix box. By this process, one obtains a surface $\Sigma$ in $V$ whose parallel copies define a fibration over $S^1$ in the complement of $\gamma$ and $\gamma'$. Namely, we have a foliation $\mathcal{F}_V$ with two screw loop defects along $\gamma$ and $\gamma'$. In addition, the layer $F$ extends as $\Sigma$ inside $V$, and thus $F$ now intersects $T^2\times \{\delta\}$ along $\Gamma_0$. This concludes the proof of the existence of $\mathcal{F}_V$, and thus the construction of the new smectic configuration $\mathcal{F}'$ with prescribed invariants. One might easily check again that if the initial configuration $\mathcal{F}$ was given by an open book, so is the new configuration $\mathcal{F}'$. In addition, in the case that $\mathcal{F}$ is given by a rational open book decomposition, every layer will intersect $T^2\times \{1\}\subset V$ along $k$ disks (which vary continuously with respect to the layer). Since the foliation $\mathcal{F}_V$ is obtained by taking parallel copies of the surfaces in each helix box, the multiplicity of each layer will be equal to $rk$. Lastly, we point out that in this construction with two helix boxes, the three new loop dislocations are the knot, a parallel copy of it, and a third loop $\gamma'$ which is not necessarily of the same knot type as the initial knot. 
\begin{Remark}
The previous considerations also show the following fact. Suppose that we fix a smectic configuration in $T^2\times [0,1]$ whose layers intersect the boundary tori $T^2\times \{0,1\}$ along parallel closed curves. Then one can replace the smectic configuration $T^2\times [0,1]$ using the surfaces given by at most two helix boxes while keeping the original boundary behavior.
\end{Remark}

 \section{Any link transverse to the layers} \label{s:link}
 
We finish this work by giving a constructive procedure by which one can deform a given link by an isotopy\footnote{Recall that we say that two links, parametrized by pairwise disjoint embeddings $\gamma_0,\gamma_1$ of $k$ copies of $S_1$ into $M$, are isotopic if there is a one-parametric family of pairwise disjoint embeddings $\gamma_t: S_1\sqcup ... \sqcup S_1 \rightarrow M$ with $t\in [0,1]$.} so that it becomes transverse to the layered structure of a smectic given by a rational open book decomposition. One can keep in mind the case in which the smectic is defined in $\mathbb{R}^3$ by the projection method introduced in \cite{SKB}. A consequence of this fact, by Section \ref{s:local}, is that on any such smectic configuration, one can locally introduce any knot or link type as a set of screw loop defects.

The method generally holds for any closed three-manifold $M$ (possibly with boundary) and a rational open book decomposition $(B,\pi)$ of $M$.  The fact that any link $L\subset M$ can be isotoped to be transverse to the pages of a standard open book can be traced back to Skora \cite{Sk}. To keep the construction explicit and self-contained, we give all the details of an argument of Becker \cite{Be} that adapts to the case of rational open book decompositions. For the sake of the exposition, we break this isotopy into two steps, which are needed in general. The first one puts the knot ``in general position" with respect to the layers. Namely, after a first isotopy, the knot will not intersect the singularities of the smectic and will have only finitely many tangency points with the regular layers. If this is already the case for the starting knot, this first step is not needed. The second and more involved step undoes the tangency points and deforms the knot so that it becomes everywhere transverse to the regular layers.

\subsection{A first isotopy: finitely many tangency points}

Fix $M$ to be an oriented closed three-manifold (the case of a manifold with boundary is analogous), a rational open book decomposition $(B,\pi)$, and an arbitrary link $L\subset M$. Up to slightly deforming $L$, we might assume that it does not intersect $B$. In particular, there is a small enough toroidal neighborhood $U$ of $B$ such that $L\cap U =\emptyset$. Here $U$ is a finite disjoint union of domains diffeomorphic to a solid torus $D^2\times S^1$, and the cores of these solid tori are the connected components of $B$. The set $M\setminus U$ is a fibration over $S^1$, with fiber a compact surface $\Sigma$ with boundary. Equivalently, we have
$$M\setminus U \cong \Sigma \times I/\sim,$$
where $\Sigma\times \{0\}$ is identified with $\Sigma \times \{1\}$ via some homeomorphism $\varphi: \Sigma \longrightarrow \Sigma$ known as the monodromy of the open book. Contrary to standard open books, the monodromy near the boundary components is not necessarily the identity, and can instead be a rotation in the circle component of the annular neighborhood of a boundary component, and even identify different boundary collar neighborhoods. However, for each boundary component, some power of the monodromy is the identity in a collar neighborhood of that component. \\

From now on, let us assume that the link $L$ has only one component and is thus a knot, that we denote by $K$. The case of several components can be treated analogously. If we parametrize $K$ by an embedding $\gamma: S^1 \longrightarrow M\setminus U$, we claim that $\gamma$ can be slightly deformed to satisfy the following property: $\gamma(t)$ is tangent to the fibers of the circle bundle only at finitely many points. There is direct proof of this fact, which is not very explicit, and thus we just give a sketch of it before giving an alternative constructive approach. The statement follows from an application of the jet transversality theorem \cite[Theorem 2.9]{Hir}. Indeed, we can consider the $2$-jet space of maps from $S^1$ to $\Sigma \times I /\sim$ (which is of dimension $10$), and the set of points that are tangent up to second order to the fibers of the circle bundle is a submanifold $A$ of dimension $8$. In particular, a generic one-dimensional manifold does not intersect $A$, since the sum of the dimensions of the two submanifolds is smaller than the dimension of the ambient space. Thus, a generic perturbation of $\gamma$ has only first-order tangencies with the fibers. Such tangencies are isolated, and thus by compactness, there must be finitely many.

We now describe an alternative constructive way to deform the link into one having finitely many tangency points. By compactness, we can subdivide $S^1$ into small enough intervals $I_k$ such that the following property is satisfied. For each $I_k$, there exists a (closed) neighborhood $V$ of $\gamma(I_k)$ in $M\setminus U$ such that $V\cap \gamma(S^1)=\gamma(I_k)$ and $\gamma(I_k)$ is an unknotted segment in $V$ (meaning that it is homotopic through embeddings in $V$ to a straight segment, relative to the endpoints). Up to refining the subdivision $I_k$, we can choose the neighborhoods $V$ to be foliated charts of the foliation given by the fibers of the surface bundle structure of $M\setminus U$. That is, we can assume that $V$ is a connected open set in $\mathbb{R}^3$ with coordinates $(x,y,z)$, in a way that $\pi|_V(x,y,z)=z$. It is now an easy task to deform $\gamma|_{I_k}$ on each $I_k$, relative to the endpoints, so that it only has finitely many tangency points to the fibers $z=\operatorname{ct}$.

\subsection{A second isotopy: undoing the tangency points} 
The second, and more involved step, is to make another isotopy to undo the tangency points and find a knot isotopic to $K$ that is everywhere positively transverse to the surface fibers. We have decomposed $M$ as $U$ and a domain diffeomorphic to a surface bundle over $S^1$ that we denote by $V=M\setminus U$. Each surface $\Sigma$ in $V\cong \Sigma \times I/\sim$ has a natural orientation, induced by the orientation of the ambient manifold $M$. If $t\in S^1$ is such that $\sigma$ is transverse to the surface fibers at $t$, we say that it is positively transverse to the fibers (which are the layers of the smectic) if the orientation induced by $\sigma'(t)$ coincides with the natural orientation of the fiber. Similarly, if the induced orientation is the opposite of the natural one, we say that $\sigma$ is negatively transverse to the fibers at $t$. Recall that in $M\setminus U$ there is an angular coordinate parametrizing the surface fibers, i.e., the layers of the smectic. Consider the set of angles
\begin{equation*}
    \{\theta_0,...,\theta_{k}\}
\end{equation*}
for which $\operatorname{Im}(\sigma)$ intersects the corresponding fiber $\pi^{-1}(\theta)$ at least at one point where $\sigma$ is tangent to the layer. We can add $0$ and $2\pi$ to this set of angles, redefine $k$ if necessary, and then assume that $\theta_0=0$, $\theta_k=2\pi$, and that $k>1$. These values give a partition of the interval $[0,2\pi]$ such that $K\cap \pi^{-1}((\theta_i,\theta_{i+1}))$ is everywhere transverse to the fibers. Such an intersection might have several connected components, though, which can be positively or negatively transverse to the fibers.

\begin{figure}[!ht]
\centering
    \begin{tikzpicture}
        \node at (0,0) {\includegraphics[scale=0.2]{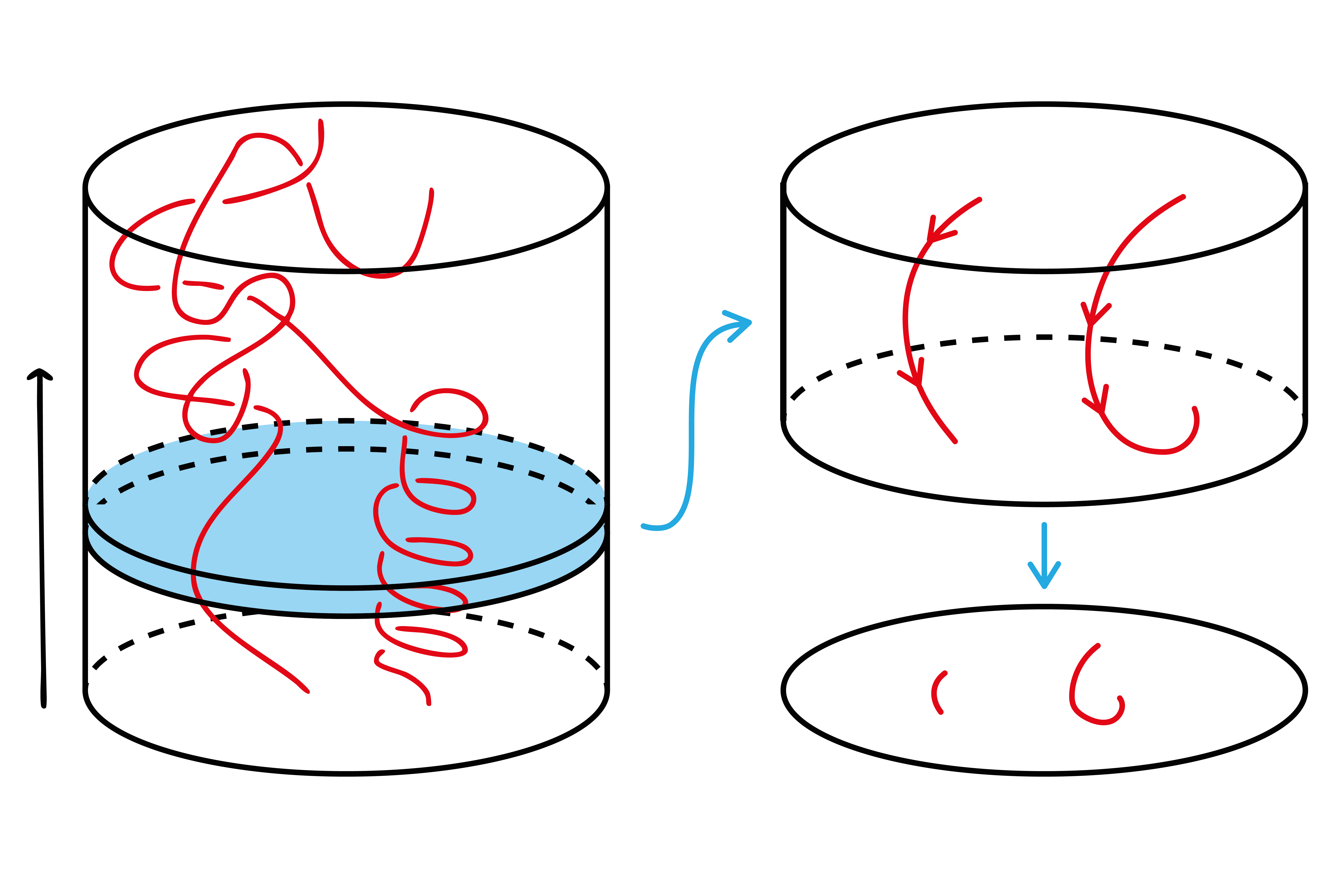}};
        \node[scale=1.2] at (-6,-0.5) {$\theta$};
        \node[scale=0.8] at (6.7, 1.1) {$\pi^{1}([\theta_i,\theta_{i+1}])$};
        \node[scale=0.9] at (3.9, -1) {$P_i$};
    \end{tikzpicture}
    \caption{Slicing the mapping torus with good projections}
    \label{fig:slicing}
\end{figure}

We claim now that we can refine this partition of the interval $[0,2\pi]$ to one that satisfies the following additional property. The set $\pi^{-1}(\theta_i, \theta_{i+1})$ is identified with $\Sigma \times (\theta_i,\theta_{i+1})$, and we want that the natural projection 
$$P_i:\Sigma \times (\theta_i,\theta_{i+1}) \longrightarrow \Sigma, $$
satisfies that $\Sigma \setminus P_i(K)$ is path-connected, see Figure \ref{fig:slicing}. Indeed, notice that $\pi^{-1}(\theta)\setminus K$ is always path-connected for any $\theta$, so for each $\theta$ and small enough $\varepsilon$, the complement of the projection of $K\cap \left(\Sigma \times (\theta-\varepsilon, \theta+ \varepsilon)\right)$ in $\Sigma$ is path-connected too. This shows that by refining the partition, we obtain a new set of angles $\theta_0,..., \theta_n$ that satisfy this additional property. For later use, we choose a connected component $C\cong S^1$ of $\partial \Sigma$ and partition it into $n+1$ disjoint intervals $I_0,...,I_{n}$.

Although this may not be important, observe that for the new partition, for some of the values $\theta_i$, the page $\pi^{-1}(\theta_i)$ might have no tangency points of $K$ with the surface fibers.\\

\begin{figure}[!ht]
    \centering
    \begin{tikzpicture}
    \node at (0,0) {\includegraphics[width=0.65\linewidth]{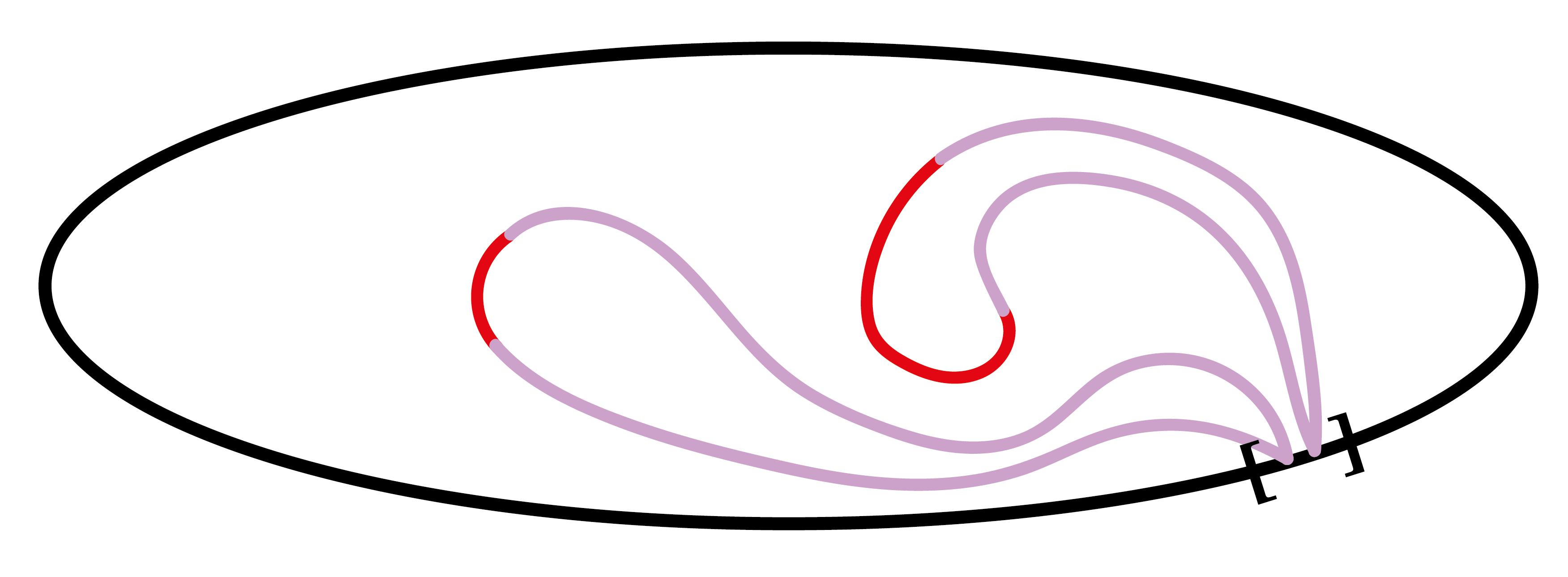}};
    \node[color=BrickRed] at (-2.2,0) {$\gamma_1$};
    \node[color=BrickRed] at (0.15, 0.5) {$\gamma_2$};
    \node at (3.15,-1.5) {$I_i$};        
    \end{tikzpicture}
    \caption{Examples of paths $\delta_j^1$ and $\delta_j^2$}
    \label{fig:paths}
\end{figure}

We will now show how to apply an isotopy to $K$ iteratively on each domain $V_i=\Sigma \times [\theta_i, \theta_{i+1}]$, perhaps by pushing $K$ inside the neighborhood $U$ of the binding, that will deform each negatively transverse part of $K$ in $V_i$ to a positively transverse one.
Fix one of the domains $V_i$ (for instance, $V_n$). Let \(K_j\), \(j = 1, \ldots, r\), be the disjoint pieces of the knot in \(V_i\). Up to renaming them, we can assume that the first \(0 \leq s \leq r\) are downwards-oriented, and the remaining \(r - s -1\) are upward-oriented. We denote by \(\gamma_j = \pi_i(K_j)\), \(j = 1, \ldots, s\) the projected curves of the downwards-oriented parts of $K$ in $V_i$. For each of them, we fix a different point $q_j$ in the interior of $I_i$ and construct two paths $\delta_j^1, \delta_j^2$ with the following properties:
\begin{itemize}
    \item[-] $\delta_j^1$ starts at $q_j$ and ends at the starting point of $\gamma_i$,
    \item[-] $\delta_j^2$ starts at the end point of $\gamma_i$ and ends at $q_j$,
    \item[-] The concatenation of paths $\delta_j^1 * \gamma_j* \delta_j^2$ bounds a disk that should not contain any other $\gamma_u$, with $u\neq j$,
    \item[-] all the paths $\{\delta_j^1, \delta_j^2\}_{j=1,...,r}$ are pairwise disjoint.
\end{itemize}
The construction of such paths is achievable thanks to the path-connectedness of \(\pi_i(V_i) \setminus \pi_i(K_i)\). Examples are depicted in purple in Figure \ref{fig:paths}.

\begin{figure}
    \centering
    \begin{tikzpicture}
        \node at (0,0) {\includegraphics[width=0.77\linewidth]{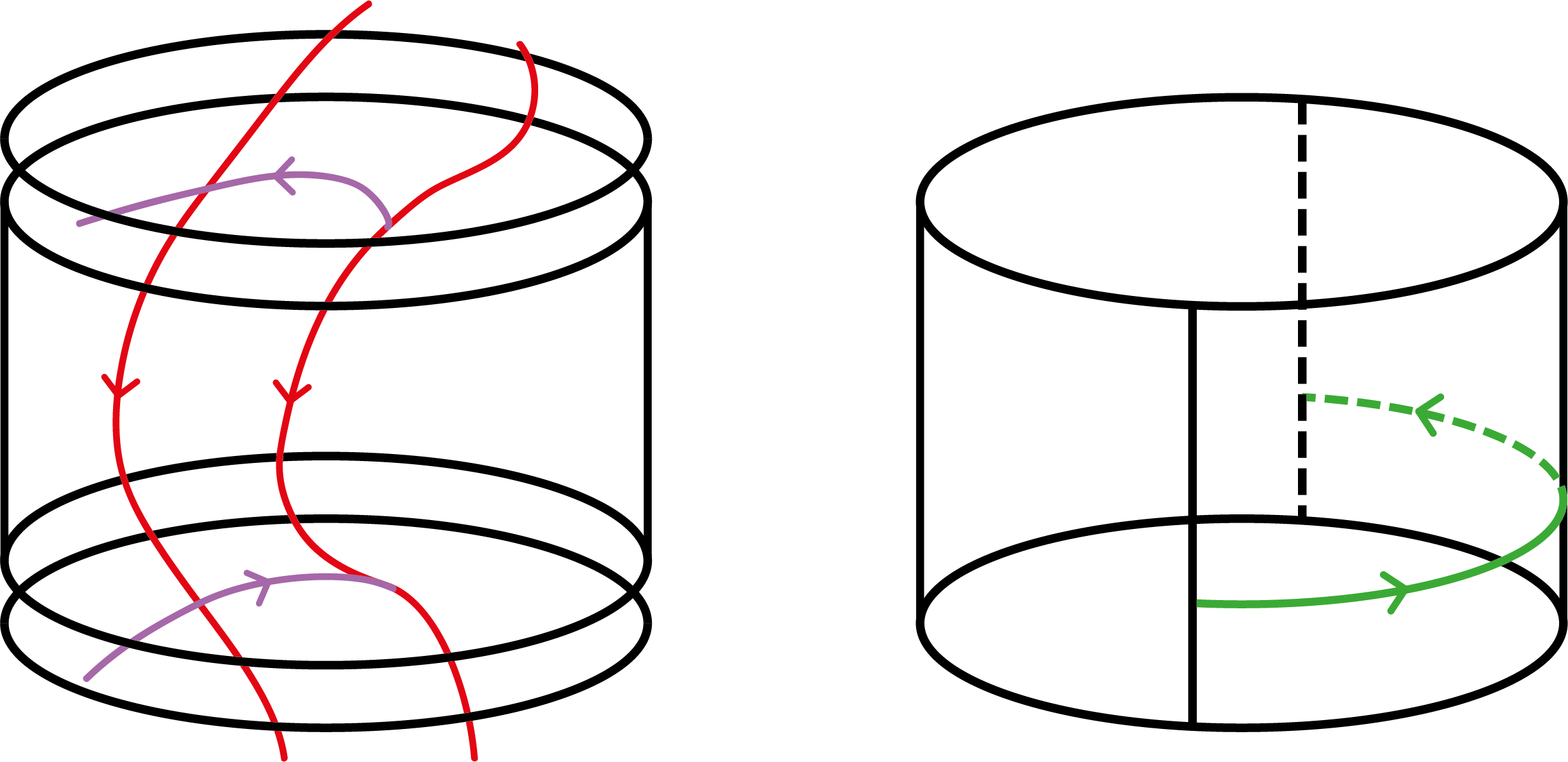}};
        \node[color=ForestGreen] at (5.75,-0.75) {$\alpha_j^3$};
        \node[color=Plum] at (-5.2, -2.45) {$\alpha_j^1$};
        \node[color=Plum] at (-5.7, 1.05) {$\alpha_j^2$};
        \node at (2.8, -2.7) {$u=u_1$};
        \node at (3.4, 2.3) {$u=u_2$};
    \end{tikzpicture}
    \caption{Constructing $\tau_j$}
    \label{fig:tau}
\end{figure}

\noindent Take $\epsilon>0$ small enough. We consider the curves $\alpha_j^{1,2}:[0,1]\rightarrow M$ given by the paths $\delta_j^{1,2}$ in the direction of the fibers, but moving as well transversely from page to page in the positive orientation. Namely, we define them by
\begin{equation*}
    \alpha_j^1(t)=(\theta_i+(t-1)\epsilon, \delta_j^1(1-t)),
\end{equation*}
\begin{equation*}
    \alpha_j^2(t)=(\theta_{i+1}+t\epsilon, \delta_j^2(t)),
\end{equation*}
which are, as we claimed, both positively transverse to the pages. In particular, if $k_j$ is a positively transverse parametrization of $K_j$, the concatenated curve $\alpha_j^1* k_j * \alpha_j^2$ is positively transverse to the pages and goes from $(\theta_{i+1}, q_j)$ to $(\theta_{i}+\varepsilon,q_j)$. To close this path, we want to define a curve that is positively transverse to the pages and goes from $(\theta_{i+1}, q_j)$ to $(\theta_i,q_j)$ along the boundary of the closure of $U$, the toroidal neighborhood of the binding. The points $(\theta_i, q_j)$ and $(\theta_{i+1}, q_j)$ both belong to the boundary $T=\partial U'$ of the closure of a connected component $U'$ of $U$ (namely, the one where the boundary component $C \subset \partial \Sigma$ is mapped). The boundary $T=\partial U'$ of $U'$ is a $2$-torus, and the fibers of the rational open book induce a foliation by curves that can be assumed to be linear. We can thus choose suitable coordinates $(u,v)$ in $\partial U\cong S^1\times S^1$ such that each page of the open book corresponds to a finite union of curves of the form $\{u=c\}$. In other words, one has that the coordinate giving the fibration of the open book satisfies $\phi=ru \mod(2\pi)$ for some integer $r$. We then construct a third path $\alpha_j^3:[0,1]\rightarrow T\subset M$ going around the binding component of $U'$ in a positively-transverse way. That is, if $(\theta_i-\varepsilon,q_j)$ corresponds to $(u_1, v_0)$ and
$(\theta_i-\varepsilon,q_j)$ corresponds to $(u_2, v_0)$, we consider the path
$$\alpha_j^3(t)= (U(t), v_0), \quad t\in [0,1]$$
were $U(t)$ turns positively from $u_1$ to $u_2$.
We have sketched these curves in an example in Figure \ref{fig:tau}. 
Finally, we consider the curve
$$\tau_{j}=k_j * \alpha_j^2 *\alpha_j^3 * \alpha_j^1.$$ 
By construction, the image of $\tau_j$ is the boundary of a set $G\subset M$ homeomorphic to a disk that does not intersect $K$. We can thus make an isotopy, relative to the endpoints, between the path $K_j$ and the image of $\alpha_j^2\circ\alpha_j^3 \circ \alpha_j^1$. This deforms $K_j$ into a path that is positively transverse to the pages, except maybe at the extreme points. At each one of the extreme points, the knot will be either positively transverse to the pages (if the extreme point was originally a tangency point), or tangent to a page (if the extreme point was originally transverse to the pages). This can be done for each \(j = 1, \dots, s\) without issues (in particular, without intersecting the knot with itself), using that the points $q_j$ that we chose in $I_i$ are all disjoint. 

\begin{figure}[!ht]
    \centering
    \begin{minipage}{1.1\linewidth} 
        \includegraphics[width=0.92\linewidth]{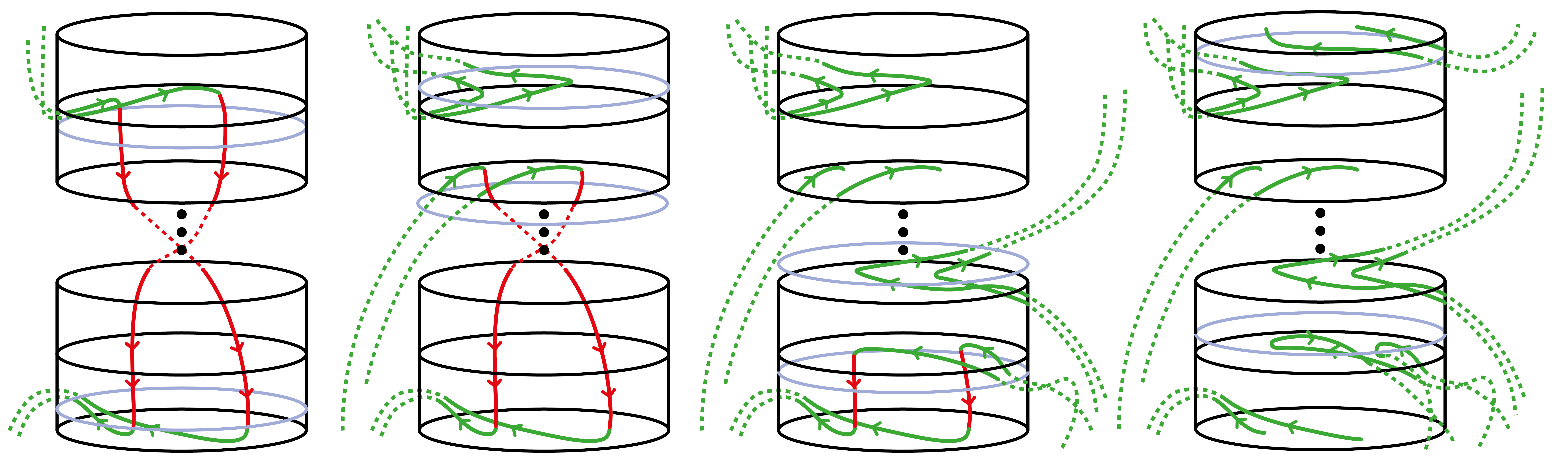}
    \end{minipage}
    \caption{Iteration over $V_i$}
    \label{fig:enter-label}
\end{figure}

In the end, notice that the pages corresponding to $\theta_i$ and $\theta_{i+1}$ will not have any more points where the knot is negatively transverse. We can now iteratively apply this procedure to $V_{i-1}$, and the isotopies we apply do not interfere with each other and do not create intersections of the knot with itself (here the fact that the intervals $I_i$ are disjoint plays a role). If we were to apply this step to $V_{i-1}$, the tangency points at $\theta_{i}$ would completely disappear. Thus, if we iterate the process down to $V_0$, at the last step, we have no tangencies left.

We have thus isotoped $K$ to be transverse to the pages of the open book as we wanted. An application of the construction described in Sections \ref{s:local} then implies that any knot type (or link type) can be locally introduced in a smectic of open book type.
\bibliographystyle{plain}
\bibliography{biblio}

\begin{thebibliography}{10}

\bibitem{Be}
Tilman Becker.
\newblock {\em On geodesible vector fields and related geometric structures}.
\newblock PhD thesis, Universit{\"a}t zu K{\"o}ln, 2023.

\bibitem{CR}
Robert Cardona and Ana Rechtman.
\newblock {Periodic orbits and Birkhoff sections of stable Hamiltonian
  structures}.
\newblock {\em Journal de l’{\'E}cole polytechnique—Math{\'e}matiques},
  12:235--286, 2025.

\bibitem{CAK}
Bryan Gin-ge Chen, Gareth~P Alexander, and Randall~D Kamien.
\newblock Symmetry breaking in smectics and surface models of their
  singularities.
\newblock {\em Proceedings of the National Academy of Sciences},
  106(37):15577--15582, 2009.

\bibitem{FB}
S{\'e}bastien Fumeron and Bertrand Berche.
\newblock Introduction to topological defects: from liquid crystals to particle
  physics.
\newblock {\em The European Physical Journal Special Topics},
  232(11):1813--1833, 2023.

\bibitem{Hat}
Allen Hatcher.
\newblock {\em Algebraic Topology}.
\newblock Cambridge University Press, Cambridge, 2002.

\bibitem{HMS}
Mikami Hirasawa, Kunio Murasugi, and Daniel~S Silver.
\newblock When does a satellite knot fiber?
\newblock {\em Hiroshima mathematical journal}, 38(3):411--423, 2008.

\bibitem{Hir}
Morris~W Hirsch.
\newblock {\em Differential topology}, volume~33.
\newblock Springer Science \& Business Media, 2012.

\bibitem{KM}
Randall~D Kamien and Ricardo~A Mosna.
\newblock The topology of dislocations in smectic liquid crystals.
\newblock {\em New Journal of Physics}, 18(5):053012, 2016.

\bibitem{M2}
Thomas Machon.
\newblock Contact topology and the structure and dynamics of cholesterics.
\newblock {\em New Journal of Physics}, 19(11):113030, 2017.

\bibitem{M1}
Thomas Machon.
\newblock The topology of knots and links in nematics.
\newblock {\em Liquid Crystals Today}, 28(3):58--67, 2019.

\bibitem{MAHK}
Thomas Machon, Hillel Aharoni, Yichen Hu, and Randall~D Kamien.
\newblock Aspects of defect topology in smectic liquid crystals.
\newblock {\em Communications in Mathematical Physics}, 372:525--542, 2019.

\bibitem{MA}
Thomas Machon and Gareth~P Alexander.
\newblock Knotted defects in nematic liquid crystals.
\newblock {\em Physical review letters}, 113(2):027801, 2014.

\bibitem{MKS}
Elisabetta~A Matsumoto, Randall~D Kamien, and Christian~D Santangelo.
\newblock Smectic pores and defect cores.
\newblock {\em Interface Focus}, 2(5):617--622, 2012.

\bibitem{Me}
N~David Mermin.
\newblock The topological theory of defects in ordered media.
\newblock {\em Reviews of Modern Physics}, 51(3):591, 1979.

\bibitem{NKTK}
Yuta Nozaki, Tam{\'a}s K{\'a}lm{\'a}n, Masakazu Teragaito, and Yuya Koda.
\newblock Homotopy classification of knotted defects in ordered media.
\newblock {\em Proceedings of the Royal Society A}, 480(2300):20240148, 2024.

\bibitem{Poe}
V~Po{\'e}naru.
\newblock Some aspects of the theory of defects of ordered media and gauge
  fields related to foliations.
\newblock {\em Communications in mathematical physics}, 80:127--136, 1981.

\bibitem{PoTh}
Joseph Pollard.
\newblock {\em The topology and geometry of liquid crystals}.
\newblock PhD thesis, University of Warwick, 2020.

\bibitem{PA1}
Joseph Pollard and Gareth~P Alexander.
\newblock Contact topology and the classification of disclination lines in
  cholesteric liquid crystals.
\newblock {\em Physical Review Letters}, 130(22):228102, 2023.

\bibitem{RG}
Andrii Repula and Eric Grelet.
\newblock Elementary edge and screw dislocations visualized at the lattice
  periodicity level in the smectic phase of colloidal rods.
\newblock {\em Physical review letters}, 121(9):097801, 2018.

\bibitem{Rol}
Dale Rolfsen.
\newblock {\em Knots and Links}.
\newblock AMS Chelsea Publishing, Providence, Rhode Island, 2003.
\newblock Reprint of 1976 edition with corrections.

\bibitem{SeKa}
Paul~G Severino and Randall~D Kamien.
\newblock Escape from the second dimension: A topological distinction between
  edge and screw dislocations.
\newblock {\em Physical Review E}, 109(1):L012701, 2024.

\bibitem{SKB}
Paul~G Severino, Randall~D Kamien, and Benjamin Bode.
\newblock Dislocations and fibrations: the topological structure of knotted
  defects in smectic liquid crystals.
\newblock {\em New Journal of Physics}, 27(6):064402, 2025.

\bibitem{Sk}
Richard~K Skora.
\newblock Closed braids in 3-manifolds.
\newblock {\em Mathematische Zeitschrift}, 211:173--187, 1992.

\bibitem{KnotsExp}
Uro{\v{s}} Tkalec, Miha Ravnik, Simon {\v{C}}opar, Slobodan {\v{Z}}umer, and
  Igor Mu{\v{s}}evi{\v{c}}.
\newblock Reconfigurable knots and links in chiral nematic colloids.
\newblock {\em Science}, 333(6038):62--65, 2011.

\bibitem{TK}
G~Toulouse and M~Kl{\'e}man.
\newblock Principles of a classification of defects in ordered media.
\newblock {\em Journal de Physique Lettres}, 37(6):149--151, 1976.

\bibitem{VH}
Jeremy Van Horn-Morris.
\newblock {\em Constructions of open book decompositions}.
\newblock The University of Texas at Austin, 2007.

\bibitem{WK}
CE~Williams and M~Kleman.
\newblock Observation of edge dislocation lines in an a-phase smectic.
\newblock {\em Journal de Physique Lettres}, 35(3):33--35, 1974.

\end{thebibliography}

\Addresses

\end{document}